\documentstyle[prd,aps,epsfig,preprint,tighten]{revtex}
\begin{document}
\preprint{\makebox{\begin{tabular}{r}
				BARI-TH/400-2000\\
				SISSA 107/2000/EP\\
\end{tabular}}}		
\title{		Analytical description of\\
		quasivacuum oscillations of solar neutrinos}
\author{	E.\ Lisi$^a$, 
		A.\ Marrone$^a$, 
		D.\ Montanino$^b$, 
		A.\ Palazzo$^a$, and
		S.T.\ Petcov$^{c,\,d,\,}$\thanks{
		Also at: Institute of Nuclear Research and
		Nuclear Energy, Bulgarian Academy of Sciences, BG--1784,
		Sofia, Bulgaria.}
}
\address{     	$^a$~Dipartimento di Fisica and Sezione INFN di Bari,\\
               	Via Amendola 173, I-70126 Bari, Italy}
\address{	$^b$~Dipartimento di Scienza dei Materiali 
		dell'Universit\`a di Lecce,\\
             	Via Arnesano, I-73100 Lecce, Italy}
\address{	$^c$~Scuola Internazionale Superiore di Studi Avanzati,\\
		Via Beirut 4, I-34014 Trieste, Italy}
\address{	$^d$~Istituto Nazionale di Fisica Nucleare, Sezione di 
		Trieste,\\ Via Valerio 2, I-34127 Trieste, Italy}
\maketitle
\begin{abstract}
%...........................................................................
We propose a simple prescription to calculate the solar neutrino survival
probability $P_{ee}$ in the  quasivacuum oscillation (QVO) regime. Such
prescription is obtained by matching  perturbative and exact analytical
results, which effectively take into account the density distribution in the
Sun as provided by the standard solar model. The resulting analytical  recipe
for the calculation of $P_{ee}$  is shown to reach its highest  accuracy
($|\Delta P_{ee}| \leq 2.6\times 10^{-2}$ in the whole QVO range)  when the
familiar prescription of choosing the solar density scale parameter $r_0$ at
the Mikheyev-Smirnov-Wolfenstein (MSW)  resonance point is replaced by a new
one, namely, when $r_0$ is chosen  at the point of ``maximal violation of
adiabaticity'' (MVA) along the neutrino  trajectory in the Sun. The MVA
prescription admits a smooth transition from the QVO regime to the MSW
transition one. We discuss in detail the phase acquired by neutrinos in the
Sun, and show that it  might be of relevance for the studies of   relatively
short timescale variations of  the fluxes of the solar $\nu$ lines  in the
future real-time  solar neutrino experiments. Finally, we elucidate the role of
matter effects in the convective zone of the Sun.
%...........................................................................
\end{abstract}

%%%%%%%%%%%%%%%%%%%%%%%%%%%%%%%%%%%%%%%%%%%%%%%%%%%%%%%%%%%%%%%%%%%%%%%%%%%%%%
\section{Introduction}
%%%%%%%%%%%%%%%%%%%%%%%%%%%%%%%%%%%%%%%%%%%%%%%%%%%%%%%%%%%%%%%%%%%%%%%%%%%%%%

The solar neutrino problem \cite{NuAs},  emerging as a  deficit of the observed
solar neutrino rates \cite{Cl98,Fu96,Ab99,Ha99,SK00,Su00,GNOC}   with respect
to standard solar model (SSM) predictions  \cite{NuAs,BP00,BaPi}, can be
explained through neutrino flavor oscillations  \cite{Po67}, possibly affected
by the presence of matter \cite{Wo78,Mi85,Ba80}  (see \cite{Revi,KrSm} for
reviews of oscillation solutions and \cite{MoPa,Vale} for recent analyses). The
analysis of neutrino oscillations requires  a detailed study of the evolution
of the  flavor states $(\nu_e,\nu_x)$, $\nu_x$  being any linear combination of
$\nu_\mu$ and $\nu_\tau$, along  the neutrino trajectory.   In the simplest
case, the active states $(\nu_e,\nu_x)$   are assumed to be superpositions,  
through a mixing angle $\omega\in[0,\pi/2]$,  of two vacuum mass eigenstates
$(\nu_1,\nu_2)$,  characterized by a mass   squared gap $\delta m^2=m^2_2-m^2_1
> 0$ (see, e.g., \cite{BiPe}).

The corresponding   neutrino evolution equation in the flavor basis,
%..............................................................................
\begin{equation}
i\frac{d}{dx}\left(
\begin{array}{c} 
\nu_e\\
\nu_x
\end{array}\right) =
H \left(
\begin{array}{c} 
\nu_e\\
\nu_x
\end{array}\right)\ ,
\label{Evol}								%Evol
\end{equation}
%..............................................................................
involves then the following vacuum $(v)$ and matter $(m)$ terms 
in the Hamiltonian $H$,
%..............................................................................
\begin{eqnarray}
H	&=& H_v + H_m \label{Hv+Hm}\\					%Hv+Hm
	&=& \frac{k}{2}\left(\begin{array}{cc}
		-\cos{2\omega} & \sin{2\omega}\\ 
		\sin{2\omega}  & \cos{2\omega}\end{array}\right)
	   +\frac{1}{2}\left(\begin{array}{cc}
		V & 0\\ 0 & -V\end{array}\right)\label{Hsplit}\ ,	%Hsplit
\end{eqnarray}
%...............................................................................
with the usual definitions for the neutrino wavenumber in vacuum,
%...............................................................................
\begin{equation}
k=\delta m^2/2E\ ,
\label{k}								%k
\end{equation}
%...............................................................................
and for the neutrino potential in matter,
%...............................................................................
\begin{equation}
V(x)=\sqrt{2}\,G_F\,N_e(x)\ ,
\label{V}								%V
\end{equation}
%...............................................................................
$N_e(x)$ being the electron density  at the point of neutrino trajectory in the
Sun, located at radial distance $x$ from the Sun center.%
%------------------------------------------------------------------
\footnote{The solar neutrinos reaching the Earth
move practically radially in the Sun. The effect of off-center trajectories 
is negligible for our results, see Appendix~A.}
%------------------------------------------------------------------
The $N_e(x)$ distribution is usually taken from SSM calculations 
\cite{BP00,BaPi}.

The Hamiltonian $H$ is  diagonalized in the matter eigenstate  basis
$(\nu_{1m},\nu_{2m})$ through  the mixing angle in matter $\omega_m$ defined by
%...............................................................................
\begin{eqnarray}
k_m\,\sin2\omega_m & = &k\,\sin2\omega\ , 
\label{sinwm}\\ 							%sinwm
k_m\,\cos2\omega_m & = &k\,\cos2\omega-V\ , 
\label{coswm}   							%coswm
\end{eqnarray}
%...............................................................................
$k_m$ being the neutrino wavenumber in matter,
%...............................................................................
\begin{equation}
k_m = k\,\sqrt{(\cos2\omega-V/k)^2+\sin^22\omega}\ .
\label{km}								%km
\end{equation}
%...............................................................................
In this work, we focus on approximate  solutions of Eq.~(\ref{Evol}) for 
neutrino flavor transitions at relatively  small values of $\delta m^2/E$
\cite{Pe87,Pe88,PeRi,Pa90,Pk91},  characterizing  the so-called   quasivacuum
oscillation (QVO) \cite{Fr00,Dark,FoQV,Spao,MoPa,Vale} regime%
%------------------------------------------------------------------------
\footnote{The lower part of the  QVO range ($\delta m^2/E\lesssim 5\times
10^{-10}$ eV$^2$/MeV) corresponds to the vacuum oscillation (VO) regime
\protect\cite{Vosc}.  In this work,  we do not explicitly distinguish the  VO
range, but simply treat  it as part of the QVO range.  Let us note that VO
solutions of the solar  $\nu$ problem are disfavored by the current data
\protect\cite{Su00,MoPa,Vale}.},
%-------------------------------------------------------------------------
%..........................................................................
\begin{equation}
{\rm QVO}\  \leftrightarrow\ \delta m^2/E \lesssim 10^{-8}
{\rm\ eV}^2/{\rm MeV} \ . 
\label{QV}
\end{equation}
%.........................................................................
In order to define our goals more precisely,  we recall that, both in the QVO
regime and in the ``Mikheyev-Smirnov-Wolfenstein'' (MSW)  \cite{Wo78,Mi85}
regime, which takes place at $\delta m^2\gtrsim ~{\rm few} \times 10^{-8}$
eV$^2$/MeV, the solar $\nu_e$ survival probability  at the Earth surface can be
written as
%...............................................................................
\begin{equation}
P_{ee}=P_0+P_1\ ,	
\label{P0P1}								%P0P1
\end{equation}
%...............................................................................
where $P_0$ is the average probability,
%...............................................................................
\begin{equation}
\textstyle
P_0=\frac{1}{2}+\left(\frac{1}{2}-P_c \right)\cos2\omega\cos2\omega_m^0\ , 
\label{P0}								%P0
\end{equation}
%...............................................................................
and $P_1$ is an oscillating 
term \cite{Pe88} (see also \cite{PeRi,Pe97}),
%...............................................................................
\begin{equation}
\textstyle
P_1=-\sqrt{P_c(1-P_c)} \cos2\omega_m^0 \sin2\omega
\cos (\Phi_{21} - \Phi_{22})\ . 
\label{P1}								%P1
\end{equation}
%..............................................................................

The main ingredients of the above  equations are: $(i)$ The  mixing angle in
matter $\omega_m^0$ at the  $\nu_e$ production point $x_0$, with
$\omega_m^0\simeq \pi/2$ in the QVO regime;   $(ii)$ The level crossing  (or
jump) probability $P_c$ of  the transition  from $\nu_{2m}$ at $x=x_0$ to
$\nu_1$ at the surface of the Sun ($x=R_\odot$); and the phase  
%...............................................................................
\begin{equation}
\textstyle
 \xi = \Phi_{21} - \Phi_{22},
\label{csphi}								%csphi
\end{equation}
%..............................................................................
accumulated on the $\nu$ path from $x_0$ to the Earth surface, $x=L=1$~A.U.,%
%-----------------------------------------------------------------------------
\footnote{In this work we consider  only  solar matter effects. Earth matter
effects are not relevant in the QVO regime,   while in the MSW regime in which
effectively $P_1 \simeq 0$ \cite{PeRi}, they can always be   implemented
through a modification of the expression for $P_{ee} \simeq P_0$, see the
discussion in  \protect\cite{FoQV}.}
%-----------------------------------------------------------------------------\.
where the phases $\Phi_{2i}$, $i=1,2$,  have a simple physical interpretation
\cite{Pe88}:  $\Phi_{2i} \equiv \Phi_{2i}(x_0,L) = {\rm arg} [A_{2i}(x_0,L)]$, 
$A_{2i}$  being the probability amplitude of  the transition  $\nu_{2m}(x_0)
\rightarrow \nu_{i}(L)$  between the initial matter-eigenstate  
$\nu_{2m}(x_0)$ and the   final vacuum-mass eigenstate  $\nu_i(L)$.  The phase
$\xi$ can also be decomposed (see, e.g., Appendix~A) as a sum   of a ``solar''
phase $\xi_s$ acquired on the path in the Sun ($x\in[x_0,R_\odot]$)  and a
``vacuum'' phase $\xi_v$, acquired in vacuum ($x\in[R_\odot,L]$),
%..........................................................................
\begin{eqnarray}
\xi&=&\xi_s + \xi_v\ 
\label{csi}\\								%csi
&=&\xi_s + k(L-R_\odot)\ .
\label{csiv}
\end{eqnarray}
%............................................................................

As a consequence of Eqs.~(\ref{P0P1})--(\ref{csiv}),  solving the neutrino
evolution  equation~(\ref{Evol}) basically reduces to calculating $P_c$ and
$\xi_s$.  This task can be accomplished through numerical integration of 
Eq.~(\ref{Evol})  for $x\in[x_0,R_\odot]$, leading to ``exact'' solutions.%
%----------------------------------------------
\footnote{In this work, high precision numerical  results (which we will call
``exact'')  for SSM density \protect\cite{BP00,BaPi} are  obtained through the
computer  codes developed in \protect\cite{FoQV}.}
%--------------------------------------------
 However, suitable approximations to the  exact results are also useful, both
to speed up the calculations and  to clarify the inherent physics.

The starting point of such approximations  is usually the analytical solution
of Eq.~(\ref{Evol}) for the case of  exponential  density $N_e(x)$ in the Sun
\cite{Pe87,Pe88} (see also \cite{To87,It88,Ab92,Hax95}).  Deviations of the SSM
density from the exponential   profile can then be  incorporated   by an
appropriate prescription for the choice of the  scale height parameter 
%...............................................................................
\begin{equation}
r_0(x) = - \left [ \frac{d}{dx} \ln N_e(x)\right ]^{-1}\ ,
\label{r0x}							%r0x
\end{equation}
%...............................................................................
characterizing the realistic change of    the electron density along the
neutrino trajectory in the Sun.  The exponential  density analytical results
are derived assuming that  $r_0 =$~const.  A well-known and  widely used 
prescription for precision calculations   of the jump probability  $P_c$ in the
MSW regime is to use the ``running''  scale height parameter  $r_0(x)$ in Eq.
(\ref{r0x}),  where for given  $\delta m^2/E$ and  for $\omega <  \pi/4$,   $x$
is chosen to coincide with the  MSW resonance point, $x = x_{\rm res}$
\cite{Kr88}.

 In the present work, we show how    the analytical solution (Sec.~II) and the
``running resonance'' prescription    for  $r_0$ (Sec.~III)  can be smoothly
extended from the MSW range down to the QVO range,  and we give a justification
for our procedure. The extension is first  achieved by matching the resonance
prescription  to a perturbative expression  for $P_c$, in the limit of small
$k$ (Sec.~III).  A more satisfactory match is  then reached (Sec.~IV) by
replacing the resonance prescription,  $r_0 = r_0(x_{\rm res})$, which can be
implemented only for $\omega<\pi/4$,  with the  ``maximal violation of
adiabaticity'' (MVA) prescription,  $r_0 = r_0 (x_{\rm mva})$, where for given 
$\delta m^2/E$ and any given  $\omega\in[0,\pi/2]$, $x_{\rm mva}$ is the point 
of the neutrino trajectory in  the Sun at which  adiabaticity is maximally
violated (or, more precisely, the adiabaticity function has a minimum).  We
also show how  the phase $\xi_s$ can  be easily and accurately calculated in 
the QVO  range through perturbative expressions  at $O(k^2)$, and discuss the 
conditions  under which $\xi_s$  might play a phenomenological role (Sec.~V).
In Sec.~VI it is shown that the perturbative results essentially probe the
low-density, convective zone of the Sun. Our final prescriptions  for the
calculation  of $P_c$ and $\xi_s$,   summarized in Sec.~VII, allow to calculate
$P_{ee}$ using Eqs. (\ref{P0P1})--(\ref{P1})  with an accuracy  $|\Delta
P_{ee}| \lesssim 2.6\times 10^{-2}$   in the whole QVO range.  All technical
details are confined in Appendixes~A--E.

While this work was being completed,   our attention was brought to the
interesting work \cite{FrQV}, where the QVO  range is  also investigated from a
viewpoint that, although being generally  different from ours, partially 
overlaps on the topic of adiabaticity  violation. We have then inserted 
appropriate comments at the end of Sec.~IV.

%%%%%%%%%%%%%%%%%%%%%%%%%%%%%%%%%%%%%%%%%%%%%%%%%%%%%%%%%%%%%%%%%%%%%%%%%%%%%
\section{Analytical forms for $P_{\lowercase{c}}$ and $\xi_{\lowercase{s}}$}
%%%%%%%%%%%%%%%%%%%%%%%%%%%%%%%%%%%%%%%%%%%%%%%%%%%%%%%%%%%%%%%%%%%%%%%%%%%%%

In this section we recall the analytical  expressions for $P_c$ and $\xi_s$,
valid for an exponential density profile,
%...............................................................................
\begin{equation}
N_e(x)=N_e(0) \exp(-x/r_0)\ ,
\label{Nexp}							%Nexp
\end{equation}
%...............................................................................
where $N_e(0)$ and $r_0$ are derived from  a best fit to the SSM profile
\cite{BP00,BaPi}  (see Fig.~1 in \cite{FoQV}),
%...............................................................................
\begin{eqnarray}
N_e(0) &= & 245 {\rm\ mol/cm}^3\ ,
\label{Ne0}\\								%Ne0
r_0 &=& R_\odot/10.54\ .
\label{r0exp}								%r0exp
\end{eqnarray}
%..............................................................................

Assuming $N_e(x)$ as in Eq.~(\ref{Nexp}), the neutrino evolution equation
(\ref{Evol}) can be solved exactly in terms of  confluent hypergeometric
functions \cite{Pe87,Pe88,To87,It88,Hax95}.  The associated expression  for
$P_c$ can  be simplified \cite{Pe87}  by making a zeroth order expansion of the
confluent hypergeometric  functions in the small parameter 
%...............................................................................
\begin{equation}
|z| \equiv r_0 V(R_\odot) \simeq 0.165\ ,
\label{small}							%small
\end{equation}
%..............................................................................
and by using the asymptotic series expansion  in the inverse powers of the
large parameter 
%..............................................................................
\begin{equation}
|z_0| \equiv r_0 V(x_0) \gtrsim 4.5 \times 10^{2}\ ,
\label{large}							%large
\end{equation}
%..............................................................................
where $x_0\lesssim 0.25\, R_\odot$ (bulk of the neutrino production zone). Then
one obtains the well-known  ``double exponential'' form for $P_c$ \cite{Pe87},
%..............................................................................
\begin{equation}
P_c=\frac{\exp(2\pi r_0 k \cos^2\omega)-1}%
{\exp(2\pi r_0 k )-1}\ .
\label{Pcanalyt}						%Pcanalyt
\end{equation}
%..............................................................................

Concerning the phase $\xi_s$,  to zeroth order in $r_0 V(R_\odot)$ one gets a
compact formula   \cite{Pe88} (without using the asymptotic series expansion
in  $|z_0|^{-n}$), 
%..............................................................................
\begin{eqnarray}
\xi_s &=&-2\arg\Gamma (1 - c) - \arg\Gamma (a - 1)
	+\arg\Gamma (a - c)\nonumber \\
	& & -r_0k\ln[r_0V(x_0)]+k(R_\odot-x_0)\ ,
\label{xianalyt}						%xianalyt
\end{eqnarray}
%..............................................................................
where 
%............................................................................
\begin{eqnarray}
 a &=& 1 + ir_0k \sin^2\omega\ ,
\label{axianalyt}\\						%axianalyt
 c &=& 1 + ir_0k\ . 
\label{cxianalyt}						%cxianalyt
\end{eqnarray}
%..............................................................................
Both Eqs.~(\ref{Pcanalyt},\ref{xianalyt}) 
are valid at any $\omega$,  including
$\omega \geq \pi/4$.%
%-----------------------------------------------------------------------------
\footnote{We remark that the restriction  $\delta m^2\cos 2\omega>0$ made at
the beginning of \protect\cite{Pe87,Pe88}, which is equivalent  to take $\omega
< \pi/4$ for the usual choice $\delta m^2>0$, was basically functional  to
obtain $P_{ee}<1/2$ in a certain region of the parameter space  (as it was
implied by the results of the Homestake experiment). However, such restriction 
does not play any role in the derivation of the Eqs.~(\protect\ref{Pcanalyt})
and  (\protect\ref{xianalyt}) in \protect\cite{Pe87,Pe88}, although this was
not emphasized at the time. This has also been recently noticed in 
\protect\cite{Mu99}.}
%-----------------------------------------------------------------------------

   Let us note that  the expression (\ref{P0P1}) for the probability  $P_{ee}$
with the average probability $P_0$ and the oscillating term $P_1$ given by Eqs.
(\ref{P0})--(\ref{csiv}), (\ref{Pcanalyt}) and   (\ref{xianalyt}) was shown
\cite{Pe88,PeRi} to assume the correct form in the VO, MSW transition   and
``large $\delta m^2$'' (averaged oscillations) regimes.

%%%%%%%%%%%%%%%%%%%%%%%%%%%%%%%%%%%%%%%%%%%%%%%%%%%%%%%%%%%%%%%%%%%%%%%%%%%%%%
\section{The modified resonance prescription }
%%%%%%%%%%%%%%%%%%%%%%%%%%%%%%%%%%%%%%%%%%%%%%%%%%%%%%%%%%%%%%%%%%%%%%%%%%%%%%

In this section, we show how the  resonance prescription for calculation of 
$P_c$ (and $P_{ee}$), valid in the MSW   regime and in the first octant of
$\omega$, can be  modified to obtain  accurate values for $P_c$ also for $k\to
0$ and for $\omega\gtrsim \pi/4$.

The resonance prescription in the MSW range   is based on the following
approximations: $(i)$ oscillations are assumed   (and where shown in
\cite{PeRi}) to be  averaged out,  so that effectively $P_1\simeq 0$ and
$\xi_s$ becomes  irrelevant%
%----------------------------------------------------------------
\footnote{The same conclusion is valid for all other oscillating terms 
(and their phases) in $P_{ee}$ \protect\cite{Pe88,PeRi}.};
%----------------------------------------------------------------- 
$(ii)$ $P_c$ is taken from Eq.~(\ref{Pcanalyt}),  but with a variable  scale
height parameter  $r_0=r_0(x)$ [Eq.~(\ref{r0x})],  with $x$  ``running'' with
the resonance  (RES) point $x=x_{\rm res}$ \cite{Kr88}, 
%............................................................................
\begin{equation}
	r_0 = r_0(x_{\rm res})\ ,
\label{r0res}							%r0res
\end{equation}
%............................................................................
where $x_{\rm res}$ is defined by the resonance  condition \cite{Mi85,Ba80}
%............................................................................
\begin{equation}
\cos 2\omega_m(x_{\rm res}) = 0\ ,
\label{omegares}						%omegares
\end{equation}
%............................................................................
when applicable. In the absence of   resonance crossing, it was customary to
take $P_c=0$ in the MSW regime (see, e.g., \cite{KuPa}).  In particular, in the
MSW analysis of  \cite{Fo96}, $P_c=0$ was taken in the second octant  of
$\omega$, where $\cos2\omega_m<0$ and Eq.~(\ref{omegares}) is never satisfied.

The indicated resonance prescription  for the calculation of $P_c$ is known to
work very well  over at least three  decades in $\delta m^2/E$  ($\sim
10^{-7}$--$10^{-4}$ eV$^2$/MeV) \cite{Kr88},  with a typical accuracy of a few 
$\times 10^{-2}$ in $P_{ee}$. However, in the lowest  MSW decade ($\delta
m^2/E  \sim 10^{-8}$--$10^{-7}$  eV$^2$/MeV), such prescription  is not very
accurate both  in the first octant \cite{Kr88}  (where it tends to
underestimate the effective value of $r_0$)  and in the second octant 
\cite{Fr00} (where $P_c$ is small but not exactly zero).  It was found
numerically  in \cite{Kr88} that, for small $k$, a relatively small  and
constant value  of $r_0$ ($\simeq R_\odot/15.4$)  provided a better 
approximation to $P_0$ in the first octant%
%---------------------------------------------------------
\footnote{The numerical results for $P_0$  were obtained in \protect\cite{Kr88}
by using the  density $N_e(x)$ provided by  the  Bahcall-Ulrich 1988 standard
solar model  \protect\cite{BU88}.}; 
%----------------------------------------------------------
this observation has been also discussed   and extended to the second octant in
\cite{Fr00}  (where $r_0\simeq R_\odot/18.4$   is used for $k\to 0$). Here we
derive (and improve) such prescriptions  by means of perturbation theory.

The key result is worked out in Appendix~B,   where we find a perturbative
solution of the neutrino evolution equation   (\ref{Evol}), in the limit of
small values $k$,  by treating the vacuum term $H_v$  as a perturbation of the
dominant matter  term $H_m$ in the Hamiltonian [Eq.~(\ref{Hsplit})].  The
$O(k)$ solution%
%-----------------------------------------------------------------
\footnote{The perturbative expansion can  be expressed in terms of
dimensionless parameters such as $kR_\odot$ or $k r_0$. However,   for
simplicity, we use the notation $O(k^n)$, $k\to 0$ etc.,  instead of
$O(k^nR_\odot^n)$, $k R_\odot\to 0$, etc.}
%-----------------------------------------------------------------
can be expressed in terms of the following dimensionless integral,
%.................................................................
\begin{equation}
I_\eta = \int_0^1 d\rho\, \exp \left(i\int_\rho^1 d\rho'V(\rho')
R_\odot\right)\ ,
\label{Ietafirst}
\end{equation}
%.................................................................
where $\rho= x/R_\odot$,  and  $V(x)$ is given by Eq.~(\ref{V}).  In
particular, the effective value of $r_0$ (in the limit of small $k$) is given
by $2\pi^{-1}R_\odot\,{\rm Im}(I_\eta)$, namely,
%.................................................................
\begin{equation}
\lim_{k\to 0} r_0 = 
\frac{2}{\pi}R_\odot 
\int_0^1 d\rho\, \sin\left(
\int_\rho^1 d\rho' \sqrt{2}\,G_F\,N_e(\rho')\,R_\odot\right)\ ,
\label{r0limit0}
\end{equation}
%.................................................................
independently of $\omega$ (i.e., both in   the first and in the second octant).

The above perturbative results  show that  the  asymptotic ($k\to 0$) effective
value for $r_0$  depends upon a   well-defined integral over the density
profile $N_e(x)$. Using the SSM profile for $N_e(x)$ \cite{BP00,BaPi},   we
find that 
%.................................................................
\begin{equation}
\lim_{k\to 0} r_0 = R_\odot/18.9\ .
\label{r0limit}
\end{equation}
%.................................................................
The same value is obtained through exact  numerical calculations.

The appearance of integrals over   the whole density profile indicates that, 
for small $k$, the behavior of $P_c$   becomes nonlocal, as was also recently
noticed in \cite{FrQV}. We further elaborate  upon the issue of nonlocality in
Sec.~VI, where we show that the $O(k)$ perturbative results are actually
dominated by matter effects in the convective zone of the Sun
($x/R_\odot\gtrsim 0.7$), where the function $N_e(x)$ resembles a power law
rather than an exponential.

 In order to match the usual resonance prescription   [$r_0=r_0(x_{\rm res})$]
with the value $r_0=R_\odot/18.9$    in the regime of small $k$, we observe
that, for the SSM density distribution \cite{BP00,BaPi} it is $r_0(x)=R_\odot
/18.9$  at  $x=0.904\, R_\odot$.  Thus, we are naturally  led to the following
``modified resonance prescription,''
%............................................................................
\begin{equation}
r_0= \left\{\begin{array}{ll} 
r_0(x_{\rm res}) & {\rm\ \  if\ } x_{\rm res}\leq 0.904\, R_\odot\ ,\\
R_\odot/18.9 & {\rm\ \  otherwise}\ ,
\end{array} 
\right.
\label{modres}							%modres
\end{equation}
%............................................................................
where ``otherwise'' includes cases   with $\omega\geq \pi/4$,  for which
$x_{\rm res}$ is not defined. Such a simple  recipe  provides a description of 
$P_c$ which is continuous  in the mass-mixing parameters,  and is reasonably
accurate  both in the QVO  range  ($\delta m^2/E\lesssim 10^{-8}$ eV$^2$/MeV) 
and in the lowest MSW decade  ($\delta m^2/E\simeq {\rm few}\times
10^{-8}$--$10^{-7}$ eV$^2$/MeV).

Figure~1 shows isolines of $P_c$ in the  bilogarithmic plane charted by the
variables%
%-----------------------------------------------------------------------
\footnote{In all the figures of this work,  we extend the  $\delta m^2/E$
interval somewhat  beyond the QVO range,  in order to display the smooth
transition to the MSW range.}
%-----------------------------------------------------------------------
$\delta m^2/E\in[10^{-10},10^{-7}]$ eV$^2$/MeV and $\tan^2\omega\in
[10^{-3},10]$.%
%-----------------------------------------------------------------------
\footnote{The variable $ \tan^2\omega$ was  introduced in
\protect\cite{Fo96} to chart both octants of the solar $\nu$ mixing angle
$\omega$ in logarithmic scale.}
%-------------------------------------------------------------------------
The solid lines refer to the exact   numerical calculation  of $P_c$, while the
dotted lines are obtained through   the analytical formula for $P_c$
[Eq.~(\ref{Pcanalyt})], supplemented with   the modified resonance
prescription  [Eq.~(\ref{modres})]. Also shown are,   in the first octant,
isolines of  resonance radius for  $x_{\rm res}/R_\odot=0.6,\,0.7,\,0.8,$  and
0.904. The maximum difference between  the exact  (numerical) results and 
those obtained using the analytic  expression for $P_c$,  Eq. (\ref{Pcanalyt}),
amounts to $|\Delta P_c|\simeq 7.5 \times 10^{-2}$,   and is typically much
smaller. Since $P_c$ is not a directly  measurable quantity,   we propagate the
results of Fig.~1 to probability amplitudes observable  at the Earth,  namely,
to the average probability  $P_0$ [Eq.~(\ref{P0})]  and to the prefactor  of
the oscillating term $P_1$ [Eq.~(\ref{P1})].%
%---------------------------
\footnote{ The phase $\xi$  is separately studied in Sec~V.}
%-----------------------------

Figure~2 shows isolines of $P_0$ for SSM density,  derived numerically (solid
lines) and by  using the analytic  expression for $P_0$ and the modified
resonance  prescription (dotted lines).  The maximum difference  is $|\Delta
P_0|\simeq 3.4\times 10^{-2}$.  Figure~3 shows, analogously, isolines of
$P_1/\cos\xi$.  The difference amounts to  $|\Delta (P_1/\cos\xi)|\lesssim
12.5\times 10^{-2}$.

From Figs.~1--3, the modified resonance prescription for $r_0$
[Eq.~(\ref{modres})] emerges as a  reasonable and remarkably simple
approximation to the exact results for $P_c$,   valid in both the MSW and the
QVO regimes. However, it does not reproduce  the exact behavior of $P_c$ with
the requisite high precision of few \% for $\delta m^2/E\sim O(10^{-8})$
eV$^2$/MeV   and $\tan^2\omega\sim O(1)$. This difference can be understood 
and removed, to a large extent, through the improved prescription discussed in
the next Section.

%%%%%%%%%%%%%%%%%%%%%%%%%%%%%%%%%%%%%%%%%%%%%%%%%%%%%%%%%%%%%%%%%%%%%%%%%%%%%%
\section{The prescription of maximum violation of adiabaticity}
%%%%%%%%%%%%%%%%%%%%%%%%%%%%%%%%%%%%%%%%%%%%%%%%%%%%%%%%%%%%%%%%%%%%%%%%%%%%%%

In this section we generalize  and improve  Eq.~(\ref{modres}),   by replacing
the point of resonance $(x_{\rm res})$   with the point  where adiabaticity is
maximally violated ($x_{\rm mva}$), more precisely, where the adiabaticity
function has its absolute minimum  on the neutrino trajectory.

Let us recall that the validity of  the resonance prescription $r_0=r_0(x_{\rm
res})$ is based on the fact that $P_c\neq 0$   in a relatively shorth part of
the $\nu$ trajectory, where the propagation  is locally nonadiabatic. The
resonance condition, however,  can be fulfilled only in the first octant of
$\omega$.  The most general condition for nonadiabaticy,  as introduced already
in the early papers \cite{Mi85,Kr88,MS87,Ba86}  on the subject, has instead no
particular restriction in $\omega$ \cite{Me86}. Such alternative  condition can
be expressed, in the basis $(\nu_{1m},\nu_{2m})$ relevant for  the calculation
of $P_c$, in terms of the ratio  between the  diagonal term $(\pm k_m/2)$ and
the  off-diagonal term ($\pm i d\omega_m/dx$) in the (traceless) Hamiltonian.
More specifically, a transition is nonadiabatic if the ratio
%............................................................................
\begin{equation}
\gamma(x)  
 \equiv \frac{k_m(x)}{|2{d\omega_m}(x)/dx|} =
\frac{k^2\sin^22\omega}{|dV(x)/dx|}
\left( 1 + \frac{(\cos2\omega - V(x)/k)^{2}}
{\sin^22\omega} \right)^{3/2}\ ,
\label{adfunction}						%adfunction
\end{equation}
%.............................................................................
satisfies the inequality  $\gamma(x) \lesssim 1$  at least in one point of the
neutrino  trajectory in the Sun. If $\gamma(x)$ is large along the whole
trajectory, $\gamma(x) \gg 1$, the transition is adiabatic.  The minimal value
of  $\gamma(x)$ identifies  the point of  ``maximum violation of
adiabaticity,'' $x_{\rm mva}$, 
%............................................................................
\begin{equation}
\gamma(x_{\rm mva}) \equiv {\rm min}\gamma(x)\ .
\label{mva0}						%mva0
\end{equation}
%.............................................................................
We show in Appendix~B, that, along the solar  $\nu$ trajectory, the MVA point 
is uniquely defined, for any $\omega$  in both octants. In particular, such
point can be unambiguously characterized through the condition%
%------------------------------------------------------------------
\footnote{A handy approximation to the MVA condition, which by-passes
the calculation of derivatives with a modest price in accuracy, is discussed
at the end of Appendix~B.}
%---------------------------------------------------------------
%............................................................................
\begin{equation}
\frac{d^2\cos2\omega_m}{dx^2}=0 {\rm\ \ \ at\ }x=x_{\rm mva}\ .
\label{mvapoint}						%mvapoint
\end{equation}
%............................................................................

In the first octant,  in general, $x_{\rm mva}$ differs from $x_{\rm res}$, 
although one can have $x_{\rm mva}\simeq x_{\rm res}$ in  some limiting cases
(see Appendix~B for a more general discussion).  For instance, as
Eq.~(\ref{adfunction}) indicates, the two points practically coincide,  $x_{\rm
mva}\simeq x_{\rm res}$,   in the case of nonadiabtic transitions  at small
mixing angles ($\sin^22\omega \ll 1$).  Indeed, let us consider for
illustration the simplified exponential case  of $r_0 = {\rm const}$. In this
(``exp'') case  it is easy to find from Eq.~(\ref{adfunction}) that:
%............................................................................
\begin{equation}
 x^{\rm exp}_{\rm res} - x^{\rm exp}_{\rm mva} = r_0 \ln
{1\over 4}\left( 1 + \sqrt{1 + 8(1 + \tan^22\omega)}\right)\ .                 
\label{xmvares}						%xmvares
\end{equation}
%..............................................................................
Obviously, at small mixing angles  ($\tan^22\omega \ll 1$) we have  $x_{\rm mva}
\simeq x_{\rm res}$. However,  this is no longer true for the nonadiabatic 
transitions at large mixing angles in the QVO regime we are interested in. In
the latter case, as it follows from Eq.~(\ref{xmvares}),  
$x_{\rm mva} < x_{\rm res}$.

We discuss in Appendix B the more realistic case of SSM density. As far as the
calculation of $P_c$ is concerned, it turns out that $x_{\rm mva}\to x_{\rm
res}$ in the limits of small $\omega$ (or of large $k$). Therefore, the MVA
condition smoothly  extends the more familiar resonance condition in both
directions of  large mixing  and of small $k$, which are relevant to pass from
the MSW to the QVO regime.  The inequality $x_{\rm mva}<x_{\rm res}$,  derived
for exponential density, persists in the QVO range for the realistic case of
SSM density, implying that 
%................................................................
\begin{equation}
r_0(x_{\rm mva}) > r_0(x_{\rm res})\ , 
\end{equation}
%................................................................
with $r_0(x)$
defined as in Eq.~(\ref{r0x}). As a conseqence, a difference arises in the
value of $P_c$ if $x_{\rm res}$ is replaced by $x_{\rm mva}$ in the
prescription for calculating $r_0$.

Explicitly, our MVA prescription reads
%............................................................................
\begin{equation}
r_0= \left\{\begin{array}{ll} 
r_0(x_{\rm mva}) & {\rm\ \  if\ } x_{\rm mva}\leq 0.904\, R_\odot\ ,\\
R_\odot/18.9 & {\rm\ \  otherwise}\ .
\end{array} 
\right.
\label{modmva}							%modres
\end{equation}
%............................................................................

Figure~4 shows curves of iso-$r_0$ obtained through Eq.~(\ref{modmva}). The
value of $r_0$ presents weak variations (within a factor of two) in the whole
mass-mixing plane and, by construction, smoothly reaches the plateau 
$r_0=R_\odot/18.9$ for $\delta m^2\lesssim{\rm few}\times 10^{-9}$ eV$^2$/MeV.

Figures~5, 6, and 7 are analogous to Figs~1, 2 and 3, respectively, modulo the
replacement of the resonance condition [Eq.~(\ref{modres})] with the MVA
condition [Eq.~(\ref{modmva})]. The MVA prescription clearly improves the
calculation of $P_c$, $P_0$, and $P_1/\cos\xi$, with an accuracy better than a
few percent in the whole plane plotted: $|\Delta P_c|\lesssim 3.7\times
10^{-2}$, $|\Delta P_0|\lesssim 2.5\times 10^{-2}$, $|\Delta
(P_1/\cos\xi)|\lesssim 3.3\times 10^{-2}$.

   We conclude that the analytical  formula for $P_c$ [Eq.~(\ref{Pcanalyt})],
valid for an exponential density, can be applied  with good accuracy to the
case of SSM density, provided that the scale  height parameter $r_0$ is chosen
according to the MVA prescription, Eq.~(\ref{modmva}).

 A final remark is in order. We agree with the author  of \cite{FrQV} about the
fact that, in order to understand better  the behavior of $P_c$ in the QVO
regime, the concept of adiabaticity  violation {\em on the whole neutrino
trajectory}  is to be preferred to the concept of  adiabaticity violation {\em
at the resonance point}.  However, we do not share his pessimism  about the
possibility of using the running  value $r_0(x_{\rm mva})$ for accurate
calculations of $P_c$: indeed, Figs.~5-7  just demonstrate this possibility. 
Such pessimism  seems to originate  from the observation that, as $k$
decreases,  $P_c$ starts to get nonlocal contributions from points rather far
from $x_{\rm mva}$  \cite{FrQV}.  In our formalism, this behavior shows up in
the $k\to 0$ limit [Eq.~(\ref{r0limit})],  where, as mentioned in the previous
Section,  the effective value of  $r_0$ gets contributions from an extended
portion of the density profile  [see also Sec.~VI and Appendix C for further
discussions].   Our prescription (\ref{modmva}),  however, effectively takes
this fact  into account, by matching the ``local'' behavior of $r_0$ for  large
$k$ [$r_0=r_0(x_{\rm mva})$] with the ``nonlocal'' behavior of $r_0$ at  small
$k$ [$r_0=R_\odot/18.9$]. In conclusion,  the MVA prescription, appropriately
modified [Eq.~(\ref{modmva})] to match the $k\to 0$ limit,  allows a
description of $P_c$ which is very accurate in the  whole QVO range,   and
which smoothly matches  the familiar resonance prescription  up in the MSW range.

%%%%%%%%%%%%%%%%%%%%%%%%%%%%%%%%%%%%%%%%%%%%%%%%%%%%%%%%%%%%%%%%%%%%%%%%%%%%%%
\section{The phase $\xi_{\lowercase{s}}$ acquired in the sun}
%%%%%%%%%%%%%%%%%%%%%%%%%%%%%%%%%%%%%%%%%%%%%%%%%%%%%%%%%%%%%%%%%%%%%%%%%%%%%%

In this Section, we discuss the last  piece for the calculation of $P_{ee}$,
namely, the solar phase $\xi_s$.  As we will see, this phase can significantly
affect  the quasivacuum oscillations of almost ``monochromatic'' solar
neutrinos (such as those associated to the $^{7}$Be and {\em pep\/}  spectra).
Indeed, there might be favorable conditions in which the phase $\xi_s$ (often
negligible in current practical calculations)   could lead to observable
effects and should thus be taken into account.

First, let us observe that, in the QVO range, the size of the solar phase
$\xi_s$ is of $O(kR_\odot)$, as indicated by Eq.~(\ref{xianalyt}) for
exponential density \cite{Pe88},  and also  confirmed through numerical
calculations for SSM density \cite{FoQV}.  Figure~9 shows, in particular, exact
results for the ratio $\xi_s/kR_\odot$, as a function of $\delta m^2/E$, using
the SSM density.   It appears that, neglecting $\xi_s$ with respect to the
vacuum phase $\xi_v$, is almost comparable to neglect $R_\odot$ as compared
with $L$. Remarkably, there are cases in which corrections of $O(R_\odot/L)$
are nonnegligible,  e.g., in the study of time variations of $P_{ee}$ over
short time scales \cite{PeRi,Pa90,Pk91,Pa91}, induced by the Earth orbit
eccentricity ($\varepsilon = 0.0167$). In fact, the fractional monthly
variation of $L(t)$ from aphelion to perihelion is   $2\varepsilon L/6 =
1.2\,R_\odot$.  Real-time experiments aiming to detect time variations of the
$\nu$ flux in monthly bins might thus test terms of $O(R_\odot/L)\sim
O(\xi_s/\xi_v)$, as also emphasized at the end of Sec.~VI in the work 
\cite{FoQV}.

Secondly, let us recall that the oscillating term  $\cos\xi$ gets averaged to
zero when the total phase $\xi$ is very large \cite{PeRi}. The approximation
$\langle \cos\xi\rangle \simeq 0$  holds in the MSW regime, but it becomes
increasingly inaccurate (and is eventually not applicable) as $k$ decreases
down to the QVO regime. In order to understand when $\xi$ starts to be
observable (at least in principle) one can consider an optimistic situation,
namely, an ideal measurement of $P_{ee}$ with a real-time detector having
perfect energy resolution, and monitoring the flux from narrowest solar $\nu$
spectra  (the Be and {\em pep\/} neutrino lines). In this case, the most 
important---and basically unavoidable---source  of smearing is the energy
integration  \cite{PeRi,BaFr,BiPo,Pa91,Cohe} over the $\nu$ lineshape.%
%-------------
\footnote{Smearing over the $\nu$ production zone is irrelevant in the QVO
regime \protect\cite{FoQV}, see also Appendix~A.}
%-------------
It has been shown in \cite{FoQV} (see also \cite{PeRi,Pk91})  that such
integrations effectively suppresses the oscillating term $P_1$ at  the Earth
through a damping factor $D$, calculable in terms of the $\nu$ lineshape. 
Figure~8 shows the factor $D$ for the {\em pep\/} line and the for two Be
lines, characterized by average energies  $\langle E\rangle = 1442$, 863.1, and
385.5 keV, respectively. The Be and {\em pep\/} lineshapes [having $O(1)$~keV
widths] have been taken from \cite{Line}  and \cite{Pa91}, respectively.
Figure~8 proves that  $\xi$ is observable, at least in principle, in the whole
QV range  $\delta m^2/E\lesssim 10^{-8}$ eV$^2$/MeV, as far as the narrowest Be
line is considered.  Of course, the observability of $\xi$ becomes more
critical (or even impossible) by increasing the initial energy spread   (e.g.,
for continuous $\nu$ spectra) or by performing measurements (like in current
experiments)  with additional and  substantial detection smearing  in the
energy or time domain \cite{PeRi}.

   With the above caveats in mind, we set out  to describe accurately $\xi_s$
in the whole QVO range. This task is accomplished  in Appendixes~D and E where,
by means of the same perturbative method applied  earlier to $P_c$, we obtain
the $O(k)$ and $O(k^2)$ expressions for $\xi_s$, respectively. The final
perturbative result is
%............................................................................
\begin{equation}
\xi_s \simeq 0.130\,(k\,R_\odot) + 1.67\times 10^{-3} 
(k\,R_\odot)^2\cos 2\omega  + O(k^3)\ ,
\label{csikk}
\end{equation}
%............................................................................
which is in excellent agreement with the exact  result for $\xi_s$ shown in
Fig.~9, in the whole QVO range.

 The $O(k)$ coefficient in Eq.~(\ref{csikk})   is just the real part of the
integral $I_\eta$ in Eq.~(\ref{Ietafirst}) (see Appendix~D)  and is  already
sufficient for an accurate description of $\xi_s$ in the QVO range. Remarkably,
the  right magnitude of this coefficient can also be obtained from the 
analytical expression (\ref{xianalyt}), which would give $\xi_s\simeq 0.116\,
kR_\odot$  (see Appendix~D). We also keep the small $O(k^2)$ term in 
Eq.~(\ref{csikk}),  because it neatly shows that $\xi_s$ starts to become
$\omega$-dependent  for increasing values of $\delta m^2/E$, consistently with
the  exact numerical results of Fig.~9.

Figure~10 shows the error one  makes on the phase $\xi_s$,  by using
increasingly better approximations,  for two representative values of
$\tan^2\omega$. The error is given as  the absolute difference between
approximate and exact results for $\xi_s$,  in units of a period ($2\pi$).  The
lowest possible approximation is simply  to neglect $R_\odot$ with respect to
$L$, so that $\xi=kL$ and $\xi_s=kR_\odot$,  namely,  neutrino oscillations are
effectively started at the Sun center (``empty Sun'').  A better approximation
is to start oscillations at the sun edge,  $\xi=k(L-R_\odot)$ and $\xi_s=0$, by
assuming that the high sun density damps  oscillations up to the surface 
\cite{Wo78,Kr85} (``superdense Sun'').  Such two approximations (and especially
the first) are widely used in phenomenological analyses. Figure~10 shows,
however, that they can produce a considerable phase shift  (even larger than a
full period for the case of ``empty'' Sun)  in the upper QVO range.   The fact
that the SSM density is  neither $N_e=0$ nor $N_e=\infty$ is correctly taken
into account through the $O(k)$ term in Eq.~(\ref{csikk}), and even more
accurately through the full $O(k^2)$ expression for $\xi_s$ in 
Eq.~(\ref{csikk}), as evident from Fig.~10.

The errors estimated in Fig.~10 show that different approximations for $\xi_s$
can affect calculations for future experiments. In particular, the use of the
familiar ``empty Sun'' or ``superdense Sun'' approximations can possibly
generate fake phase shifts in time variation analyses for   real-time detectors
sensitive to neutrino lines, such as Borexino \cite{BORE} or KamLAND
\cite{KAML}. According to the estimates in \cite{Fr99}, such two experiments
might be sensitive to monthly-binned seasonal variations for $\delta
m^2\lesssim 5\times 10^{-9}$. Figure~10 shows that, in the upper part of such
sensitivity range, the empty Sun approximation gives $\xi_s$ (and thus $\xi$)
totally out of phase, as compared with exact results. Analogously,  the
superdense Sun approximation induces a phase shift that, although smaller than
in the previous case, can still be as large as $\pi/2$  $(\Delta \xi_s\lesssim
0.2 \times 2\pi \sim \pi/2)$ in the quoted  sensitivity range, and can thus
produce a big difference [$\Delta \cos\xi \simeq O(1)$] in the calculation of
$P_1$.

It is an unfortunate circumstance, however,  that the dominant term in $\xi_s$
is proportional to $k$. This fact implies that, neglecting  $\xi_s\simeq 0.130
R_\odot/L$ in the total phase $\xi=\xi_v+\xi_s$, is basically equivalent to 
introduce a small bias of the kind $\delta m^2\to\delta m^2\times
(1+0.130\,R_\odot/L)= \delta m^2\times (1+6\times 10^{-4})$. On one hand,  such
bias is sufficient to produce a substantial difference in  $\cos\xi$ when
$\delta m^2/E $ approaches $10^{-8}$ eV$^2$/MeV, and is thus observable in
principle.  On the other hand, $\delta m^2$ is not known a priori, but must be
derived from the experiments themselves, and it will be hardly known with a
precision of $O({\rm few}\times 10^{-4})$ for some time. Therefore, although
$\xi_s$ may produce a big effect at {\em fixed\/} values of $\delta m^2$, it
might be practically obscured by uncertainties in the {\em fitted\/} value of
$\delta m^2$.

In conclusion, we have found a simple  and accurate expression for the solar
phase $\xi_s$, to be used in  the calculation of the total phase 
$\xi=\xi_s+k(L-R_\odot)$. The solar  phase $\xi_s$ produces effects of
$O(R_\odot/L)$, which can be nonnegligible  in high-statistics, real-time
experiments sensitive to short-time variations associated to  neutrino lines,
such as Borexino and KamLAND, as was  also emphasized in \cite{PeRi,Pa90} and
more recently in \cite{FoQV}.   In such context, we recommend the use  of
Eq.~(\ref{csikk}) for precise  calculations of $P_{ee}$ at fixed value of
$\delta m^2$,  although the   observability of $\xi_s$ certainly represents a
formidable challenge.

%%%%%%%%%%%%%%%%%%%%%%%%%%%%%%%%%%%%%%%%%%%%%%%%%%%%%%%%%%%%%%%%%%%%%%%%%%%%%
\section{Probing the convective zone of the sun}
%%%%%%%%%%%%%%%%%%%%%%%%%%%%%%%%%%%%%%%%%%%%%%%%%%%%%%%%%%%%%%%%%%%%%%%%%%%%%

In this section we elaborate upon the $O(k)$ perturbative results discussed
previously for $P_c$ and $\xi_s$ (and detailed in Appendix~C and D). In
particular, we show that they are related to the density in the convective zone
of the Sun, corresponding to $\rho=x/R_\odot\gtrsim 0.7$.

The $O(k)$ results crucially depend upon the quantity $I_\eta$, which is
defined in Eq.~(\ref{Ietafirst}) as an integral over an oscillating function,
having the density $N_e(x)$ as inner argument. Numerical evaluation of such
integral for SSM density gives $I_\eta=(13.03 + i\,8.32)\times 10^{-2}$ (see
Appendix~C). It turns out that the largest contribution to $I_\eta$ comes from
the outer regions of the Sun, where the integrand oscillates slowly, while in
the inner regions the integrand oscillates rapidly, with vanishing net
contribution to the real and imaginary parts of $I_\eta$. Numerical inspection
shows that the value of $I_\eta$ is dominated by the $\rho\gtrsim 0.7$ range,
which happens to correspond to the convective zone of the Sun  \cite{NuAs}.
Therefore, it is sufficient to consider  such zone to estimate $I_\eta$.

In the convective zone, the density profile $N_e(x)$ is better described by a
power law rather than by an exponential  function (see, e.g., \cite{BaWe}). A
good approximation to the SSM density for $\rho \gtrsim 0.7$ is:
%.....................................................................
\begin{equation}
N_e(\rho)\simeq N_p (1-\rho)^{p-1}\ ,
\label{Np}					%Np
\end{equation}
%...................................................................
with $p\simeq 2.8$ and $N_p\simeq 1.4$~mol/cm$^3$. By adopting such expression
for the density, and using the position $z=1-\rho$, we get the following
expression for $I_\eta$,
%......................................................................
\begin{equation}
I_\eta = \int^1_0 dz \, \exp(i qz^p)\ ,
\label{ietaconv}				%ietaconv
\end{equation}
%........................................................................
where $q=V_p R_\odot/p \simeq 135$ and $V_p=\sqrt{2}G_F N_p$. Once again, we
note that the above integrand gives a very small contribution outside the
convective zone $(z\gtrsim 0.3)$, since the exponent becomes large. Therefore,
the upper limit can be shifted from $1$ to $\infty$ without appreciable
numerical changes, and with the advantage that the result can be cast in a
compact analytical form, reproducing the exact SSM numerical result 
with good accuracy: 
%......................................................................
\begin{eqnarray}
I_\eta &\simeq& \int^\infty _0 dz \, \exp(i qz^p)
\label{ietaconv2}\\				%ietaconv2
&=& q^{-1/p} \Gamma(1+p^{-1})e^{i\frac{\pi}{2 p}}
\label{ietaconv3}\\				%ietaconv3
&\simeq & (13 +i\, 8.2)\times 10^{-2}\ .
\label{ietaconv4}				%ietaconv4
\end{eqnarray}
%........................................................................

Using the above equation and the results of Appendixes~C and D,
the small-$k$ limit for $r_0$ and $\xi_s$ can be explicited as
%........................................................................
\begin{eqnarray}
\frac{r_0}{R_\odot} &\simeq & \frac{2}{\pi}\, q^{-1/p}\,\Gamma(1+p^{-1})\,
\sin(\pi/2p)\ ,\\
\frac{\xi_s}{kR_\odot} &\simeq & q^{-1/p}\,\Gamma(1+p^{-1})\,
\cos(\pi/2p)\ .
\end{eqnarray}
%........................................................................
Such results show that the effective  values of $r_0$ and of  $\xi_s$  at small
$k$ are connected to the parameters $(p,q)$ describing the power-law dependence
of $N_e(x)$ in the  convective zone of the Sun [Eq.~(\ref{Np})]. Therefore,
while for relatively large $k$   neutrino oscillations in matter probe  the
inner ``exponential bulk'' of the solar density profile,  for small $k$  they
mainly probe the outer,  ``power-law'' zone of convection.

Finally, let us notice that, in the range  $0.7\lesssim \rho\lesssim 0.9$,
where the ``running'' MVA prescription  $r_0=r_0(x_{\rm mva})$ is applicable,
the power law in Eq.~(\ref{Np}) leads  to $r_0(x)/R_\odot =  (1-\rho)/(p-1)$,
so that  $r_0(x)/R_\odot\simeq 1/18.9$ at the matching point $\rho\simeq
0.904$, consistently  with the prescription in Eq.~(\ref{modmva}).

%%%%%%%%%%%%%%%%%%%%%%%%%%%%%%%%%%%%%%%%%%%%%%%%%%%%%%%%%%%%%%%%%%%%%%%%%%%%%%
\section{Summary of recipes for calculation of 
$P_{\lowercase{ee}}$ in the QVO regime}
%%%%%%%%%%%%%%%%%%%%%%%%%%%%%%%%%%%%%%%%%%%%%%%%%%%%%%%%%%%%%%%%%%%%%%%%%%%%%%

We summarize our best recipe for calculation  of $P_{ee}$ as follows. In the
QVO regime, for any given value of the  mixing angle $\omega$  and of the
neutrino wavenumber $k=\delta m^2 /2E$, the $\nu_e$ survival  probability
reads 
%............................................................................
$$
P_{ee} = P_c\cos^2\omega  + (1-P_c)\sin^2\omega + \sin2\omega\sqrt
{P_c(1-P_c)}\cos\xi\ ,
$$
%...........................................................................
where
%............................................................................
$$
\xi = \xi_s + k(L-R_\odot)\ .
$$
%...........................................................................

The value of $P_c$ can be evaluated with high precision by using the analytical
form 
%............................................................................
$$
P_c=\frac{\exp(2\pi r_0 k \cos^2\omega)-1}%
{\exp(2\pi r_0 k )-1}\ ,
$$
%.............................................................................
provided that the density scale parameter $r_0$ is calculated as
%............................................................................
$$
r_0= \left\{\begin{array}{ll} 
r_0(x_{\rm mva}) & {\rm\ \  if\ } x_{\rm mva}\leq 0.904\, R_\odot\ ,\\
R_\odot/18.9 & {\rm\ \  otherwise}\ ,
\end{array} 
\right.
$$
%............................................................................
where $x_{\rm mva}$ is the point where  adiabaticity is maximally violated
along the neutrino trajectory in the Sun (see Appendix B). Such expression for
$P_c$  smoothly joins the more  familiar resonance prescription when passing
from the QVO to  the MSW regime.

 The value of the Sun phase $\xi_s$  can be calculated with high precision by
using the   $O(k^2)$ perturbative result, valid for  $\delta m^2/E \lesssim 
10^{-8}$ eV$^2$/MeV, 
%....................................................................
$$
\xi_s \simeq 0.130\,(k\,R_\odot) + 1.67\times 10^{-3} 
(k\,R_\odot)^2\cos 2\omega\ .
$$
%...................................................................
We have shown that such phase can play a role  in precise calculations of
$P_{ee}$ in the QVO range. It is not  necessary to extend the calculation of
$\xi_s$ in the MSW range, where effectively   $\langle\cos\xi\simeq 0\rangle$,
and $\xi_s$ is not observable even in principle. Notice that the  neutrino
production  point $x_0$ does not appear in the calculation of both $P_c$ and
$\xi_s$ in the QVO range.

 Finally, we have checked that the above  recipe allows the calculation of
$P_{ee}$ with an accuracy  $|\Delta P_{ee}|\lesssim 2.6 \times 10^{-2}$ (and
often much better than a percent) in  both octants of $\omega$ for the whole
QVO range ($\delta m^2\lesssim 10^{-8}$ eV$^2$/MeV).   Figure~11 shows, as an
example, a graphical comparison with  the exact results for $P_{ee}$ at the
exit from the Sun ($x=R_\odot$).%
%----------
\footnote{The function $P_{ee}$ at the Earth ($x=L$) can not be usefully
plotted in the range of Fig.~11, due to its rapidly oscillating behavior. }
%----------
It appears at glance  that our recipe represents an accurate  substitute to
exact numerical calculations of $P_{ee}$, in the whole QVO  range $\delta
m^2/E\lesssim 10^{-8}$ eV$^2$/MeV.

%%%%%%%%%%%%%%%%%%%%%%%%%%%%%%%%%%%%%%%%%%%%%%%%%%%%%%%%%%%%%%%%%%%%%%%%%%%%%%
\section{Conclusions}
%%%%%%%%%%%%%%%%%%%%%%%%%%%%%%%%%%%%%%%%%%%%%%%%%%%%%%%%%%%%%%%%%%%%%%%%%%%%%%

We have worked out a simple and accurate   prescription to calculate the solar
neutrino survival probability $P_{ee}$ in  the  quasivacuum oscillation
regime.  Such prescription adapts the known analytical solution for the
exponential case to the true  density case, as well as to the perturbative
solution of neutrino  evolution equations in the limit of small $\delta m^2/E$.
The accuracy of the  prescription is significantly improved (up to $2.6\times
10^{-2}$ in $P_{ee}$ in the  whole QVO range) by replacing the  familiar
prescription of choosing  the scale height parameter $r_0$ at the MSW resonance
point by a new one:  $r_0$ is chosen  at the point of maximal violation of
adiabaticity (MVA) along the neutrino  trajectory in the Sun.  Such
generalization preserves a smooth transition of our prescription from the QVO
to the MSW oscillation regime, where the two prescriptions practically
coincide.  We show that at sufficiently small  $k = \delta m^2/2E$ in the QVO
regime, the effective value of  $r_0$ is determined by an integral over the 
electron density distribution in the Sun, $N_e(x)$:
%................................................................
$$
\lim_{k\to 0} r_0 = \frac{2}{\pi} R_\odot 
\int_0^1 d\rho\, \sin\left(
\int_\rho^1 d\rho'\,\sqrt{2}\,G_F\, N_e(\rho')\,R_\odot\right)\ ,
\label{r0limit00}
$$
%.................................................................
where $\rho= x/R_\odot$. The main contribution in the above integral is shown 
to come from the region corresponding to  the convective zone of the Sun, $x
\gtrsim 0.7R_\odot$. Thus, if quasivacuum oscillations take place, solar
neutrino experiments might provide information about the  density distribution
in the convective zone of the Sun.  We also discuss in detail the phase 
acquired by neutrinos in the Sun,  whose observability, although possible in
principle, poses a formidable  challenge for future experiments aiming to
observe short timescale  variations of fluxes from solar $\nu$ lines.

%%%%%%%%%%%%%%%%%%%%%%%%%%%%%%%%%%%%%%%%%%%%%%%%%%%%%%%%%%%%%%%%%%%%%%%%%%%%
\acknowledgments
A.M.\ acknowledges kind hospitality at SISSA during the initial stage of this
work. S.T.P.\ would like to acknowledge the hospitality of the Aspen Center for
Physics  where part of this work was done. The work of  E.L., A.M., and A.P.\
was supported in part by the Italian  MURST under the program ``Fisica
Astroparticellare.'' The work of S.T.P.\  was supported in part  by the EEC
grant ERBFMRXCT960090 and by the   Italian MURST under the program ``Fisica
Teorica delle Interazioni Fondamentali.''

%%%%%%%%%%%%%%%%%%%%%%%%%%%%%%%%%%%%%%%%%%%%%%%%%%%%%%%%%%%%%%%%%%%%%%%%%%%%%
\appendix\section{General form of $P_{\lowercase{ee}}$ and 
its QV limit}
%%%%%%%%%%%%%%%%%%%%%%%%%%%%%%%%%%%%%%%%%%%%%%%%%%%%%%%%%%%%%%%%%%%%%%%%%%%%%

In this Appendix we recall the derivation of the basic equations for $P_{ee}$
given in the Introduction, together with further details relevant for
Appendixes B--E.

Given an initial solar $\nu_e$ state $\Psi_e=(1,0)^T$ ($T$ being the
transpose), its final survival amplitude $A_{ee}$ at the detector  can be
factorized as
%...............................................................................
\begin{eqnarray}
A_{ee}(x_0,L) &=& 
	\Psi_e^T\, U_{\omega}\,U_v\,U_m\,U^T_{\omega^0_m}\, \Psi_e
\label{AeeU}\\								%AeeU
       &=&
	\Psi_e^T
	\left(\begin{array}{cc}
	\cos\omega & \sin\omega \\ -\sin\omega & \cos\omega
	\end{array}\right) 
	\left(\begin{array}{cc}
	e^{i\xi_v/2} & 0 \\ 0 & e^{-i\xi_v/2}
	\end{array}\right)\times\nonumber\\
       & &
	\left(\begin{array}{cc}
	\sqrt{1-P_c}\, e^{-i\alpha}& -\sqrt{P_c}\,e^{-i\beta}\\ 
	\sqrt{P_c}\,e^{i\beta} & \sqrt{1-P_c}\,e^{i\alpha}
	\end{array}\right) 
	\left(\begin{array}{cc}
	\cos\omega_m^0 & -\sin\omega_m^0 \\ \sin\omega_m^0 & \cos\omega_m^0
	\end{array}\right)
	\Psi_e \ ,
\label{Aee}								%Aee
\end{eqnarray}
%...............................................................................
where the $U$ matrices act as follows (from right to left): (i)
$U^T_{\omega_m^0}$  rotates (at $x=x_0$) the initial state $N_e$  into the
matter mass basis; (ii) $U_m$ is a generic $SU(2)$ parametrization for the
evolution of the matter mass eigenstates  $(\nu_{1m},\nu_{2m})$ from the origin
($x=x_0$) up to the exit from the Sun $(x=R_\odot)$ where $\nu_{im}=\nu_i$, in
terms of the so-called crossing probability $P_c=P(\nu_{2m}\to\nu_{1})$  and of
two generic phases $\alpha$ and $\beta$; $(iii)$ $U_v$ evolves the mass
eigenstates in vacuum along one astronomical distance $L$, with $\xi_v$ defined
in Eq.~(\ref{csiv});  and $(iv)$  $U_\omega$ finally rotates the mass basis
back to the flavor basis at the detection point ($x=L$).%
%-------------
\footnote{During nighttime, one should insert a fifth  matrix to take into
account the evolution within the Earth, which we do not consider in this work
(focussed on solar matter effects in the QVO range).   Earth matter effects can
always be added afterwards as a calculable modification to $P_0$,   since they
turn out to be nonnegligible only in the MSW regime, when oscillations are
averaged out and $P_1\simeq 0$,  see \cite{FoQV} and references therein. }
%-----------

After some algebra,  the resulting $\nu_e$ survival probability can be
expressed as the sum of an ``average term'' $P_0$ plus four ``oscillating
terms'' \cite{Pe88,Pe97},
%...............................................................................
\begin{equation}
P_{ee}=|A_{ee}(x_0,L)|^2=\sum_{n=0}^4 P_n\ ,
\label{Pee}
\end{equation}
%...............................................................................
where
%...............................................................................
\begin{eqnarray}
P_0 &=& \frac{1}{2}+
	\left(\frac{1}{2}-P_c\right)\cos2\omega_m^0\cos2\omega
\ ,\label{AP0}\\							%AP0
P_1 &=& -\cos2\omega_m^0\sin2\omega\sqrt{P_c(1-P_c)}\cos(\xi_v+\pi-\alpha-\beta)
\ ,\label{AP1}\\							%AP1
P_2 &=& -\sin2\omega_m^0\cos2\omega \sqrt{P_c(1-P_c)}\cos(\alpha-\beta) 
\ ,\label{AP2}\\							%AP2
P_3 &=& - \frac{1}{2}\sin2\omega_m^0\sin2\omega P_c 
	[\cos(\xi-2\alpha)+\cos(\xi_v-2\beta)]
\ ,\label{AP3}\\							%AP3
P_4 &=& \frac{1}{2}\sin2\omega_m^0\sin2\omega\cos(\xi_v-2\alpha)
\ .\label{AP4}							%AP4
\end{eqnarray}
%...............................................................................
It has been shown in \cite{Pe88,PeRi} 
that the last three terms $P_{2,3,4}$ can, in general,
be safely neglected for practical purposes, so that one can take 
%...............................................................................
\begin{equation}
P_{ee} = P_0+P_1\ ,
\label{AP0P1}								%AP0P1
\end{equation}
%..............................................................................
which, together with the identification
%...............................................................................
\begin{equation}
\xi_s = \pi-\beta-\alpha\ ,
\label{xiodot}							%xiodot
\end{equation}
%...............................................................................
leads to Eqs.~(\ref{P0P1})--(\ref{P1}).%
%-----------------------------
\footnote{One can make contact with the $\Phi_{ij}$ phase notation of 
\protect\cite{Pe88,PeRi,Pe97} through the following identifications:
$\Phi_{11}=-\alpha+\xi_v/2$;  $\Phi_{12}=\beta-\xi_v/2$;
$\Phi_{21}=\pi-\beta+\xi_v/2$; and $\Phi_{22}=\alpha-\xi_v/2$. The notation in
the present work  explicitly factorizes out the contribution of the vacuum
phase  $\xi_v$ for $x\in [R_\odot,L]$. }
%-----------------------------

In the QVO regime relevant for  our work ($k\lesssim 10^{-8}$ eV$^2$/MeV), the
negligibility  of $P_{2,3,4}$ is evident  from the fact that 
$k/V(x_0)\lesssim  4\times 10^{-3}$  for $x_0\lesssim 0.25 R_\odot$, so that
the  $P_{2,3,4}$ prefactor $\sin 2\omega_m^0$ is negligibly small, while the
nonvanishing terms $P_{0,1}$ can be written as:
%.............................................................................
\begin{eqnarray}
P_0^{\rm QVO} &\simeq & \cos^2\omega P_c + \sin^2\omega (1-P_c)\ , 
\label{P0QV}\\							%P0QV
P_1^{\rm QVO} &\simeq & 2\sin\omega\cos\omega\sqrt{P_c(1-P_c)}\cos\xi\ .
\label{P1QV}							%P1QV
\end{eqnarray}
%.............................................................................
It was shown in \cite{FoQV}  that the above equations are not spoiled by Earth
matter effects,%
%-----------------------------------------------------------------------
\footnote{Equations~(\protect\ref{P0QV}) and (\protect\ref{P1QV}) in this work 
coincide with Eq.~(28) in \cite{FoQV}, modulo  the identification  $P_\odot =
P_c$, valid in the QVO regime.}
%-----------------------------------------------------------------------
as they turn out to be negligible in the QVO range---a fortunate circumstance
that considerably simplifies the calculations.

We have  verified the applicability of the approximations in Eq.~(\ref{P0QV})
and (\ref{P1QV}) in the QVO regime,  by checking that our exact results do not
change  in any appreciable way by  setting $\cos\omega_m^0 \equiv 0$ from the
start. In particular, the results  of the numerical integration of the
evolution equations vary very little by forcing the initial state to be
$\nu_{2m}$ rather than $\nu_e$. We have also verified that none of our figures
changes in a graphically perceptible way  in the QV range, by moving the point
$x_0$ within the production zone $(x_0\lesssim 0.25 R_\odot)$ not only radially
but also for off-center trajectories.%
%---------------
\footnote{ Notice that the $x_0$-independence of  $P_{ee}^{\rm QVO}$ implies
that no smearing over the production zone is necessary, as also emphasized in
\protect\cite{FoQV}.}
%---------------------
Therefore, in the following appendixes, we will neglect corrections of
$O(k/V(x_0))$ in the QVO regime, and  just set 
%...........................................................................
\begin{eqnarray}
x_0 &\equiv& 0\ ,
\label{x00}\\ 							%x00
\omega_m^0 &\equiv & \pi/2\ ,
\label{pi/2}							%pi/2
\end{eqnarray}
%.............................................................................
from the start,  without any appreciable loss of accuracy.

Given Eq.~(\ref{Aee}), and the fact that the initial state  can be taken, to a
high degree of accuracy, $\nu_e = \nu_{2m}$,  the probability amplitudes to
find the two mass eigenstates at $x=R_\odot$ ($\xi_v=0$) simply read
%...........................................................................
\begin{eqnarray}
A_{21} &=& \sqrt{P_c}e^{i(\pi-\beta)}\ ,
\label{nu1m} \\							%nu1m
A_{22} &=& \sqrt{1-P_c}e^{i\alpha}\ .
\label{nu2m} 							%nu2m
\end{eqnarray}
%.............................................................................

In terms of flavor state transition amplitudes $A_{ee}$ 
and $A_{ex}$   at $x=R_\odot$, the jump probability reads
%.............................................................................
\begin{eqnarray}
P_c^{\rm QVO} &=& |A_{21}|^2\ , % |\nu_{1}|^2\ 
\\
&=& |A_{ee}|^2\cos^2\omega+|A_{ex}|^2\sin^2\omega-
2\,{\rm Re}(A_{ee}\, A_{ex}^*)\sin\omega\cos\omega\ ,
\label{PcQV} 							%PcQV
\end{eqnarray}
%..............................................................................
and the phase reads%
%-----------------------------------------------------------------------
\footnote{Equations~(\protect\ref{PcQV}) and 
(\protect\ref{csiQV}) can also be obtained  from
Eqs. (13a)--(13d), (15), and (19) in \protect\cite{Pe97}.} 
%-----------------------------------------------------------------------
%..............................................................................
\begin{eqnarray}
\xi_s^{\rm QVO} &=& \arg(A_{21}\, A^*_{22}) %\arg(\nu_{1}\,\nu^*_{2})
\\
&=& \arg[\sin\omega \cos\omega(|A_{ee}|^2-|A_{ex}|^2)
+{\rm Re}(A_{ee} A_{ex}^*)\cos2\omega
+i\,{\rm Im}(A_{ee} A_{ex}^*)]\ .
\label{csiQV} 							%csiQV
\end{eqnarray}
%.............................................................................

We conclude this Appendix by quoting  
useful expressions for some dimensionless
quantities which appear in the calculation of $P_c$ and $\xi$:
%..............................................................................
\begin{eqnarray}
L/R_\odot &=& 215\ ,
\label{LRs}\\								%L/Rs	
k R_\odot &=& 1.76 \times 10^9
\left( \frac{\delta m^2/E}{{\rm eV}^2/{\rm MeV}}\right)\ ,
\label{kRs}\\								%kRs
V R_\odot &=& 2.69 \times 10^2
\left( \frac{N_e}{{\rm mol/cm}^3}\right)\ ,
\label{VRs}\\								%VRs
V/k&=&1.53 \times 10^{-7} 
		\left(\frac{\delta m^2/E}{{\rm eV}^2/{\rm MeV}}\right)^{-1}
		\left(\frac{N_e}{{\rm mol}/{\rm cm}^3}\right)\ .
\label{V/k}								%V/k
\end{eqnarray}
%...............................................................................

%%%%%%%%%%%%%%%%%%%%%%%%%%%%%%%%%%%%%%%%%%%%%%%%%%%%%%%%%%%%%%%%%%%%%%%%%%%%%
\section{The condition of maximal violation of adiabaticity}
%%%%%%%%%%%%%%%%%%%%%%%%%%%%%%%%%%%%%%%%%%%%%%%%%%%%%%%%%%%%%%%%%%%%%%%%%%%%%

In this Appendix we characterize  the condition of maximum violation of
adiabaticity along the neutrino trajectory.  We also show that the MVA
condition reduces to the familiar resonance  condition in appropriate limits.

In the matter eigenstate basis, the neutrino evolution equation
%..............................................................................
\begin{equation}
i\frac{d}{dx}\left(
\begin{array}{c} 
\nu_{1m}\\
\nu_{2m}
\end{array}\right) =
H' \left(
\begin{array}{c} 
\nu_{1m}\\
\nu_{2m}
\end{array}\right)\ ,
\label{Evolmatter}						%Evolmatter
\end{equation}
%..............................................................................
is governed by the Hamiltonian
%...............................................................................
\begin{equation}
H'=\frac{k_m}{2}\left(
\begin{array}{cc}
-1 & -2\,i\,{\dot\omega_m}/{k_m} \\
2\,i\,{\dot\omega_m}/{k_m} & 1
\end{array}
\right)\ .
\label{Hmatter}							%Hmatter
\end{equation}
%...............................................................................
where $\dot\omega_m=d\omega_m/dx$.

The evolution is adiabatic when the diagonal term dominates over the
off-diagonal term, so that the ratio of such elements [the function $\gamma(x)$
defined in Eq.~(\ref{adfunction})] is large. The ``maximal violation of 
adiabaticity'' along the neutrino trajectory is reached at the point $x=x_{\rm
mva}$ where $\gamma(x)$ is instead minimized, 
%...............................................................................
\begin{equation}
\frac{d\gamma(x)}{dx}=0\ \leftrightarrow\ x=x_{\rm mva}\ .
\label{mva1}							%mva1
\end{equation}
%...............................................................................
The above MVA condition can be rewritten as
%...............................................................................
\begin{equation}
\frac{d^2\cos 2\omega_m}{dx^2} =0\ \leftrightarrow\ x=x_{\rm mva}\ . 
\label{mva2}							%mva2
\end{equation}
%...............................................................................
Since $\cos 2\omega_m$ increases monotonically from its value at the production
point ($\simeq -1$) to its vacuum value  $(\cos 2\omega)$, the above condition
characterizes the (flex) point of fastest increase for the function
$\cos2\omega_m(x)$, which is thus uniquely defined at any $\omega\in[0,\pi/2]$.

The MVA condition (\ref{mva2}) can be compared with the usual resonance 
condition,
%...............................................................................
\begin{equation}
\cos 2\omega_m =0\ \leftrightarrow\ x=x_{\rm res}\ . 
\label{res}							%mva2
\end{equation}
%...............................................................................
which can be fulfilled only for $\omega\in[0,\pi/4]$. To understand under which
circumstances the two conditions practically coincide ($x_{\rm mva}\simeq
x_{\rm res}$), it is useful to rewrite the MVA condition in the following,
equivalent form:
%...............................................................................
\begin{equation}
3\sin2\omega_m\cos 2\omega_m(\dot V)^2+k 
\sin2\omega \ddot V =0\ \leftrightarrow\ x=x_{\rm mva}\ . 
\label{mva3}							%mva2
\end{equation}
%...............................................................................
The above condition reduces  to Eq.~(\ref{res}) in the case  linear density
$(\ddot V=0)$ (which, however,  is not applicable to the Sun). It also reduces
to Eq.~(\ref{res}) for very  small mixing ($\sin 2\omega\to 0$), which explains
why the simple  resonance condition  has been very frequently used in the
literature, to replace the slightly  more complicated MVA condition. Notice
also that, for increasing $k$,  $P_c$ becomes nonnegligible  only at small
mixing angles. Therefore, as far as the calculation  of $P_c$ is concerned, the
MVA and  resonance conditions also merge  at large $k$. In all other cases, it
turns out that $x_{\rm res}>x_{\rm mva}$ for $\omega\in [0,\pi/4]$.  For
$\omega\in [\pi/4,\pi/2]$, $x_{\rm res}$ is simply not defined.

We conclude by providing a handy approximation to the value of $x_{\rm mva}$,
which may be used to by-pass the calculation of derivatives of $V(x)$, at the
price of a small loss of accuracy in the calculation of $P_{ee}$. As discussed
before, the MVA condition characterizes the flex point of the curve
$\cos\omega_m(x)$, which increases from $-1$ to $\cos 2\omega$ in the QVO
range.  We have verified that, in the parameter region where $P_c$ is
nonnegligible, the flex point $x_{\rm mva}$ is close to the half-rise point
$x_{1/2}$ where $\cos2\omega_m$ assumes the average value between its two
extrema, defined as
%...................................................................
\begin{equation}
\cos 2\omega_m(x_{1/2}) \equiv \frac{1}{2}(\cos2\omega-1)\ .
\label{x12}						%lable{x12}
\end{equation}
%..................................................................
The above definition leads to the condition
%..................................................................
\begin{equation}
V(x_{1/2})=k\cos 2\omega + \frac{k\sin^3\omega}{\sqrt{1+\sin^2\omega}}\ ,
\end{equation}
%..................................................................
which clearly reduces to the resonance condition in the small $\omega$ limit.
Using $r_0(x_{1/2})$ as a substitute for $r_0(x_{\rm mva})$ in the MVA
prescription [Eq.~(\ref{modmva})] provides a final accuracy of $\lesssim
4.5\times 10^{-2}$ in $P_{ee}$ in the whole QV regime  (only slightly worse
than the accuracy quoted at the end of Sec.~VII).

%%%%%%%%%%%%%%%%%%%%%%%%%%%%%%%%%%%%%%%%%%%%%%%%%%%%%%%%%%%%%%%%%%%%%%%%%%%%%
\section{Expansion of $P_{\lowercase{c}}$ at first order in $\lowercase{k}$}
%%%%%%%%%%%%%%%%%%%%%%%%%%%%%%%%%%%%%%%%%%%%%%%%%%%%%%%%%%%%%%%%%%%%%%%%%%%%%

In this Appendix we study the  limit $k\to 0$ for  $P_c$, in both cases of 
exponential and SSM density. Such  limit gives an  asymptotic value for the
effective scale parameter $r_0$,  relevant for the lower part of the QV regime.
The positions in  Eqs.~(\ref{x00})  and (\ref{pi/2}) are adopted. 
We also introduce a normalized radial coordinate,
%...............................................................................
\begin{equation}
\rho = x/R_\odot\ .
\label{rho}							%\rho 
\end{equation}
%...............................................................................

In the case of exponential density, Eq.~(\ref{Pcanalyt}) gives, at first order
in $k$, 
%...............................................................................
\begin{equation}
P_c = \cos^2\omega -\frac{\pi}{4}\,r_0\, k \sin^2 2\omega\ ,
\label{Pck}							%Pck
\end{equation}
%...............................................................................
where $r_0 = R_\odot/10.54$ [Eq.~(\ref{r0exp})].

In the case of SSM density,  we demonstrate that Eq.~(\ref{Pck}) is formally
preserved, modulo the replacement
%...............................................................................
\begin{equation}
r_0 \to r'_0 = R_\odot/18.9\ .			
\label{r0'}							%r0'
\end{equation}
%...............................................................................
The crucial observation is that, for $k\lesssim 10^{-9}$ eV$^2$/MeV and SSM
density, we have $k\ll V(x)$ for most of the $\nu$ trajectory inside the Sun
($\rho \lesssim 0.9$) before the rapid density drop at the border.  This
suggest  a perturbative solution of Eqs.~(\ref{Evol}) and (\ref{Hv+Hm}) with
the vacuum term $H_v$ playing the role of a small perturbation, as compared
with the dominant matter term $H_m$. The perturbative expansion is expected to
be accurate when the second term in Eq.~(\ref{Pck}) 
is small, namely, for $kr_0 \ll 1$.

At 0th order in $H_v$, the evolution operator from  $\rho=0$ to $\rho=1$ is
trivial in the flavor basis ($H_m$ being diagonal),
%...............................................................................
\begin{eqnarray}
T_0(1,0)&=&\exp \left(-i\int^1_0 d\rho \, H_m(\rho)\right) 
\label{T0}\\							%T0
&=& \left(\begin{array}{cc}
\exp\left[-\frac{i}{2}\eta(1,0)\right] & 0 \\
0 & \exp\left[+\frac{i}{2}\eta(1,0)\right]  
\end{array}\right)\ ,
\label{T0'}							%T0'
\end{eqnarray}
%...............................................................................
where
%...............................................................................
\begin{equation}
{\eta(\rho_2,\rho_1)=\int^{\rho_2}_{\rho_1}d\rho\,V(\rho)R_\odot}\ .
\label{eta}							%eta
\end{equation}
%...............................................................................

At 1st order in $H_v$ (i.e., in $k$), one gets an improved evolution operator
$T_1$,
%...............................................................................
\begin{equation}
T_1(1,0) = T_0(1,0) [{\mathbf{1}}-i\Delta(1,0)]\ , 
\label{T1}							%T1
\end{equation}
%...............................................................................
where the correction matrix $\Delta$ reads
%...............................................................................
\begin{eqnarray}
\Delta(1,0) &=& T_0^{-1}(1,0)
\int^1_0 d\rho\, T_0(1,\rho)\,H_v\,T_0(\rho,0) 
\label{Delta}\\							%Delta
&=& \frac{kR_\odot}{2}
\left(\begin{array}{cc}
-\cos2\omega & \sin2\omega \displaystyle\int_0^1 d\rho\, e^{i\eta(\rho,0)}\\
\sin2\omega \displaystyle\int_0^1d\rho\, e^{-i\eta(\rho,0)} & \cos2\omega 
\end{array}
\right)\ .
\label{Delta'}							%Delta'
\end{eqnarray}
%...............................................................................

By applying $T_1$ to the initial flavor state $\psi_e=(1,0)^T$, one gets
the flavor state at $x=R_\odot$,
%...............................................................................
\begin{equation}
\left(\begin{array}{c}\nu_e\\ \nu_x\end{array}\right)_{R_\odot}
= e^{-\frac{i}{2}\eta(1,0)}
\left[
\left(\begin{array}{c}1 \\ 0 \end{array}\right)
-\frac{i}{2}kR_\odot
\left(\begin{array}{c}-\cos2\omega \\ 
\sin2\omega\displaystyle \int^1_0d\rho\, e^{i\eta(1,\rho)} \end{array}\right)
\right]\ ,
\label{flavor}							%flavor
\end{equation}
%...............................................................................
where the $O(k)$ correction is explicited. The basic integral,
%...............................................................................
\begin{equation}
I_\eta = \int_0^1 d\rho \exp[i\,\eta (1,\rho)] = C+iS\ , 
\label{Ieta}							%Ieta
\end{equation}
%...............................................................................
appearing in Eq.~(\ref{flavor}), can be numerically evaluated using the SSM
density,
%...............................................................................
\begin{eqnarray}
C &=&\int_0^1 d\rho\, \cos\int_\rho^1 d\rho'V(\rho')R_\odot = 0.1303\ ,
\label{C}\\							%C
S &=& \int_0^1 d\rho\, \sin\int_\rho^1 d\rho'V(\rho')R_\odot = 0.0832\ .
\label{S}							%S
\end{eqnarray}
%...............................................................................

Finally,  using Eqs.~(\ref{PcQV},\ref{flavor})  and keeping only $O(k)$ terms,
one gets
%...............................................................................
\begin{eqnarray}
P_c 
&\simeq & \cos^2\omega - \sin2\omega \, {\rm Re }(\nu_e\nu_x^*)\\
&\simeq & \cos^2\omega -\frac {1}{2}k R_\odot S \sin^2 2\omega\ ,
\end{eqnarray}
%...............................................................................
which reproduces Eq.~(\ref{Pck}), up to the anticipated replacement
%...............................................................................
\begin{equation}
r_0\to r'_0= 2\,S\, R_\odot/\pi =R_\odot/18.9\ .
\label{replacer0}						%replacer0
\end{equation}
%...............................................................................
We have independently checked the value $r_0'=R_\odot/18.9$ by numerical
studies of the $k\to 0$ limit  for the quantity $(P_c-\cos^2\omega)/k$, with
$P_c$ derived exactly rather than perturbatively.%
%-----
\footnote{The author of 
\protect\cite{Fr00,FrQV} found numerically a similar value for $r_0'$
(namely, $R_\odot/18.4$), by  fitting his exact results for small $k$. In this
Appendix we have given an explicit derivation of $r_0'$, based on perturbation
theory, and independent on numerical fits. }
%-------

We observe that, for the SSM density, the scale parameter $r_0(x)$  happens to
be equal to the value in Eq.~(\ref{r0'}) at $x=0.904 \,R_\odot$,
%...............................................................................
\begin{equation}
r_0(0.904\, R_\odot) = R_\odot/18.9 = r'_0\ .
\label{r09}
\end{equation}
%...............................................................................
Such relation proves useful to match our analytical and perturbative results
around 
the critical zone of the ``knee'' of the SSM density, which occurs just at
$x\simeq 0.9 R_\odot$.

As a final exercise, we prove that $r'_0 \simeq r_0$ in the exponential limit
$V(x)=V(0)e^{-x/r_0}$, as it should be. The proof uses the assumptions of 
large initial density and small final density,  as discussed for
Eqs.~(\ref{small}) and (\ref{large}). Then one changes  the integration
variable from $\rho$ to $\phi=V(0)r_0e^{-\rho R_\odot/r_0}$ in the calculation
of $S$ [Eq.~(\ref{S})], and exploits the fact that $\phi$ is very small (large)
for $\rho=1$ ($\rho=0$):
%...............................................................................
\begin{eqnarray}
\frac{r'_0}{R_\odot} &=& \frac{2}{\pi}\int^1_0 d\rho\,
\sin\int_\rho^1 d\rho'\, V(0)\,R_\odot\, 
e^{-\rho' R_\odot/r_0} \\
&\simeq & \frac{2}{\pi} \int^1_0 d\rho\,\sin\left(V(0)\,r_0\,e^{-\rho 
R_\odot/r_0}\right)\\
&\simeq& 
\frac{r_0}{R_\odot}\frac{2}{\pi}\int^\infty_0 d\phi\,\frac{\sin\phi}{\phi}
= \frac{r_0}{R_\odot}\ .
\label{boh}
\end{eqnarray}
%...............................................................................

%%%%%%%%%%%%%%%%%%%%%%%%%%%%%%%%%%%%%%%%%%%%%%%%%%%%%%%%%%%%%%%%%%%%%%%%%%%%%
\section{Expansion of $\xi_{\lowercase{s}}$ at first order in $\lowercase{k}$}
%%%%%%%%%%%%%%%%%%%%%%%%%%%%%%%%%%%%%%%%%%%%%%%%%%%%%%%%%%%%%%%%%%%%%%%%%%%%%

In this Appendix we study the small-$k$ limit of the phase $\xi_s$, by using
the same strategy adopted in the previous section for $P_c$. In particular, let
us consider the $O(k)$ expansion of the analytical phase in
Eq.~(\ref{xianalyt}),
%...............................................................................
\begin{equation}
\xi_s \simeq k\,R_\odot\, C'\ ,
\label{csik}\\						%csik
\end{equation}
%..............................................................................
where%
%---------
\footnote{$\gamma_E\simeq0.577$ is the Euler constant.}
%---------
%..............................................................................
\begin{equation}
C'=1-\frac{r_0}{R_\odot}[\gamma_E+\ln r_0V(0)] =0.116\ .
\label{C'}						%C'
\end{equation}
%...............................................................................

We show that Eq.~(\ref{csik}), valid for exponential density, remains formally
unchanged also for  the SSM density, modulo the replacement $C'\to C$, where
$C=0.130$ is given by Eq.~(\ref{C}). In fact, using Eqs.~(\ref{csiQV}) and
(\ref{flavor}), and keeping only the $O(k)$ terms, we get
%...............................................................................
\begin{eqnarray}
\xi_s
&\simeq & 2 {\rm Im}(A_{ee}A_{ex}^*)/\sin 2\omega\\
&=& k\,R_\odot\, C\ .
\label{csi0k}
\end{eqnarray}
%...............................................................................
Such result is independently confirmed by studying the $k\to 0$ limit of the
quantity $\xi_s/kR_\odot$, with $\xi_s$ derived from exact (rather than perturbative)
calculations.

Finally, we show that $C\simeq C'$ in the limit of exponential density, as it
should be. With manipulations similar to those used to derive Eq.~(\ref{boh}) 
we get:
%......................................................................
\begin{eqnarray}
C &=& \int_0^1 d\rho \,\cos\int^1_\rho d\rho'\,V(0)\,R_\odot
\,e^{-\rho'R_\odot/r_0}\\
&\simeq& \int^1_0 d\rho\,\cos\left(V(0)\,r_0\,e^{-\rho 
R_\odot/r_0}\right)\\
&=& 1-\int^1_0 d\rho\,\left[1-\cos\left(V(0)\,r_0\,e^{-\rho 
R_\odot/r_0}\right)\right]\\
&=& 1-\frac{r_0}{R_\odot}\int^{r_0 V(0)}_0\frac{1-\cos\phi}{\phi}
\label{diverg}\\						%diverg
&\simeq& 1-\frac{r_0}{R_\odot}[\gamma_E+\ln r_0V(0)] = C'
\end{eqnarray}
%........................................................................
where, in the divergent integral of Eq.~(\ref{diverg}), we have kept only the
two leading  terms [$\gamma_E$ and $\ln r_0V(0)$] which do not decrease when
the upper limit $r_0V(0)$ becomes large.

%%%%%%%%%%%%%%%%%%%%%%%%%%%%%%%%%%%%%%%%%%%%%%%%%%%%%%%%%%%%%%%%%%%%%%%%%%%%%
\section{Expansion of $\xi_{\lowercase{s}}$ at second order in $\lowercase{k}$}
%%%%%%%%%%%%%%%%%%%%%%%%%%%%%%%%%%%%%%%%%%%%%%%%%%%%%%%%%%%%%%%%%%%%%%%%%%%%%

In this Appendix, we present an $O(k^2)$ perturbative expression for $\xi_s$,
which effectively accounts for the $\omega$-dependence of the exact results 
(as evident in the upper part of the QVO range in Fig.~9).

We basically iterate the calculation of the evolution operator $T(1,0)$ (see
Appendix~C) at second order in the perturbation term $H_v$. We omit the
(somewhat lengthy) derivation and present the final $O(k^2)$ result,
%...............................................................................
\begin{eqnarray}
\xi_s &=&  (k R_\odot)\, {\rm Re}I_\eta +  (k R_\odot)^2\,\cos2\omega\,
[{\rm Im}(I_\eta-I'_\eta+I''_\eta)/2-{\rm Re}I_\eta \cdot {\rm
Im}I_\eta]
 \label{csi2k}\\						%csi2k
&=& 0.130\, (k R_\odot) + 1.67 \times 10^{-3} (k R_\odot)^2\,\cos 2\omega\ ,
\label{csi2k'}							%csi2k'
\end{eqnarray}
%...............................................................................
where $I_\eta = (13.03 + i\,8.32)\times 10^{-2}$ is defined 
in Eq.~(\ref{Ieta}), while $I'_\eta$ and $I''_\eta$ are defined as 
%...............................................................................
\begin{eqnarray}
I'_\eta &=& \int^1_0 d\rho\, \rho\,\exp[i\,\eta(1,\rho)] \\ 
&=&
(12.45 + i\, 7.07)\times 10^{-2}\ ,
\label{Ieta'}\\
I''_\eta &=& \int^1_0 d\rho \int^\rho_0 d\rho' \,
 \exp[i\,\eta(1,\rho')]
\\ &=&  (0.58 + i\, 1.25)\times 10^{-2}\ ,
\label{Ieta''}
\end{eqnarray}
%...............................................................................
and have been  numerically evaluated for SSM density. The $O(k)$ term
reproduces, of course, the results given in the previous section.

The $O(k^2)$ behavior of Eq.~(\ref{csi2k'})  is not captured by the analytical
phase written in the form of  Eq.~(\ref{xianalyt}), which contains only odd
powers in a $k$-expansion. The reason can be traced to the fact that
Eq.~(\ref{xianalyt}) was derived in  \cite{Pe88,Pe97} under the assumption of
zero density at $x=R_\odot$ (which is not exactly fulfilled for exponential
density).  Preliminary studies indicate that removal of such assumption gives 
the correct $O(k^2)$ behavior of the analytical phase [including the
$\cos2\omega$ dependence displayed in Eq.~(\ref{csi2k'})]. Such studies are
proceeding and will be presented after completion.

%%%%%%%%%%%%%%%%%%%%%%%%%%%%%%%%%%%%%%%%%%%%%%%%%%%%%%%%%%%%%%%%%%%%%%%%%%%%%%%
% 			R E F E R E N C E S 
%%%%%%%%%%%%%%%%%%%%%%%%%%%%%%%%%%%%%%%%%%%%%%%%%%%%%%%%%%%%%%%%%%%%%%%%%%%%%%%

%\end{document}

%%%%%%%%%%%%%%%%%%%%%%%%%%%%%%%%%%%%%%%%%%%%%%%%%%%%%%%%%%%%%%%%%%%%%%%%%%%%
%%%%%%%%%%%%%%%%%%%%%%%%%%%%%%%%%%%%%%%%%%%%%%%%%%%%%%%%%%%%%%%%%%%%%%%%%%%%
%%%%%%%
%%%%%%%            INCLUSION OF FIGURES WITH EPSFIG.STY.
%%%%%%%
%%%%%%%%%%%%%%%%%%%%%%%%%%%%%%%%%%%%%%%%%%%%%%%%%%%%%%%%%%%%%%%%%%%%%%%%%%%%
%%%%%%%          P O S T S C R I P T       F I G U R E S 
%%%%%%%
%%%%%%%   memo:  1) add epsfig in the \documentstyle
%%%%%%%          2) and move this part before \end{document} 
%%%%%%%		 3) remove previous figure captions
%%%%%%%          4) include the following \newcommand:
%%----------------------------------------------------------------------------
\newcommand{\InsertFigure}[2]{\newpage\phantom{.}
\vspace*{-2.cm}\begin{center}\mbox{%
\epsfig{bbllx=2truecm,bblly=2truecm,bburx=19.5truecm,bbury=26.5truecm,%
height=23.truecm,figure=#1}}\end{center}\vspace*{-2.truecm}%
\parbox[t]{\hsize}{\small\baselineskip=0.5truecm\hskip0.5truecm #2}}
%----------------------------------------------------------------------------
%%----------------------------------------------------------------------------
\newcommand{\ShrinkFigure}[2]{\newpage\phantom{.}
\vspace*{-0.cm}\begin{center}\mbox{%
\epsfig{bbllx=2truecm,bblly=2truecm,bburx=19.5truecm,bbury=26.5truecm,%
height=20.truecm,figure=#1}}\end{center}\vspace*{-1.truecm}%
\parbox[t]{\hsize}{\small\baselineskip=0.5truecm\hskip0.5truecm #2}}
%----------------------------------------------------------------------------
%%%%%%%%%%%%%%%%%%%%%%%%%%%%%%%%%%%%%%%%%%%%%%%%%%%%%%%%%%%%%%%%%%%%%%%%%%%%%%%

%..............................................................................
\InsertFigure{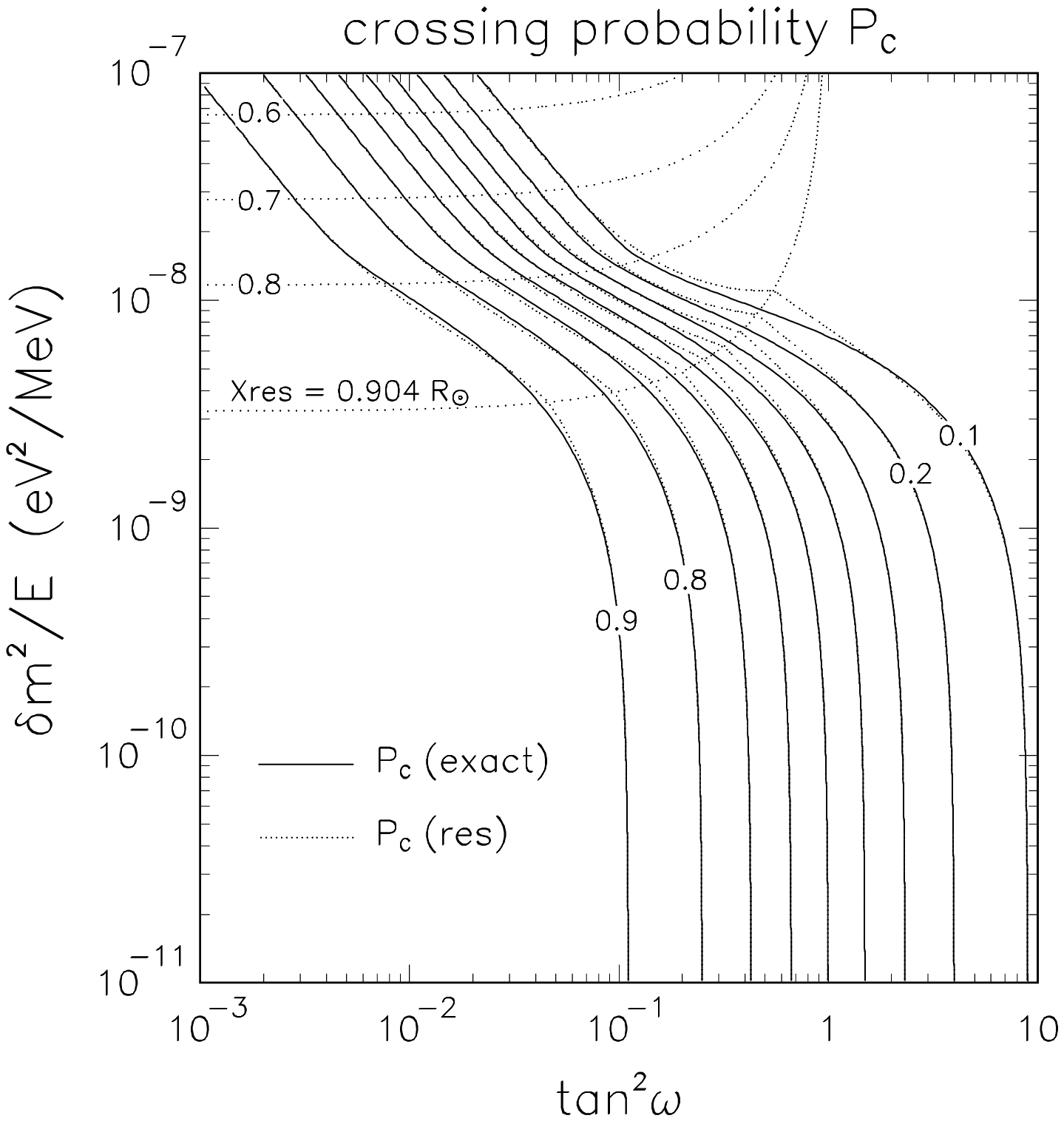}%
{FIG.~1. The crossing probability $P_c$, calculated for the SSM density in the
parameter space $(\delta m^2/E,\,\tan^2\omega)$. The three lowest decades in
$\delta m^2/E$ characterize the quasivacuum oscillation regime. Solid curves:
exact numerical results. Dotted curves: approximate results, using the
analytical resonance prescription.  Also shown are isolines of resonance radii.
The radius $x/R_\odot=0.904$ represents the matching point of the resonance
prescription with the perturbative results valid for small values of $\delta
m^2/E$.}
%..............................................................................
\InsertFigure{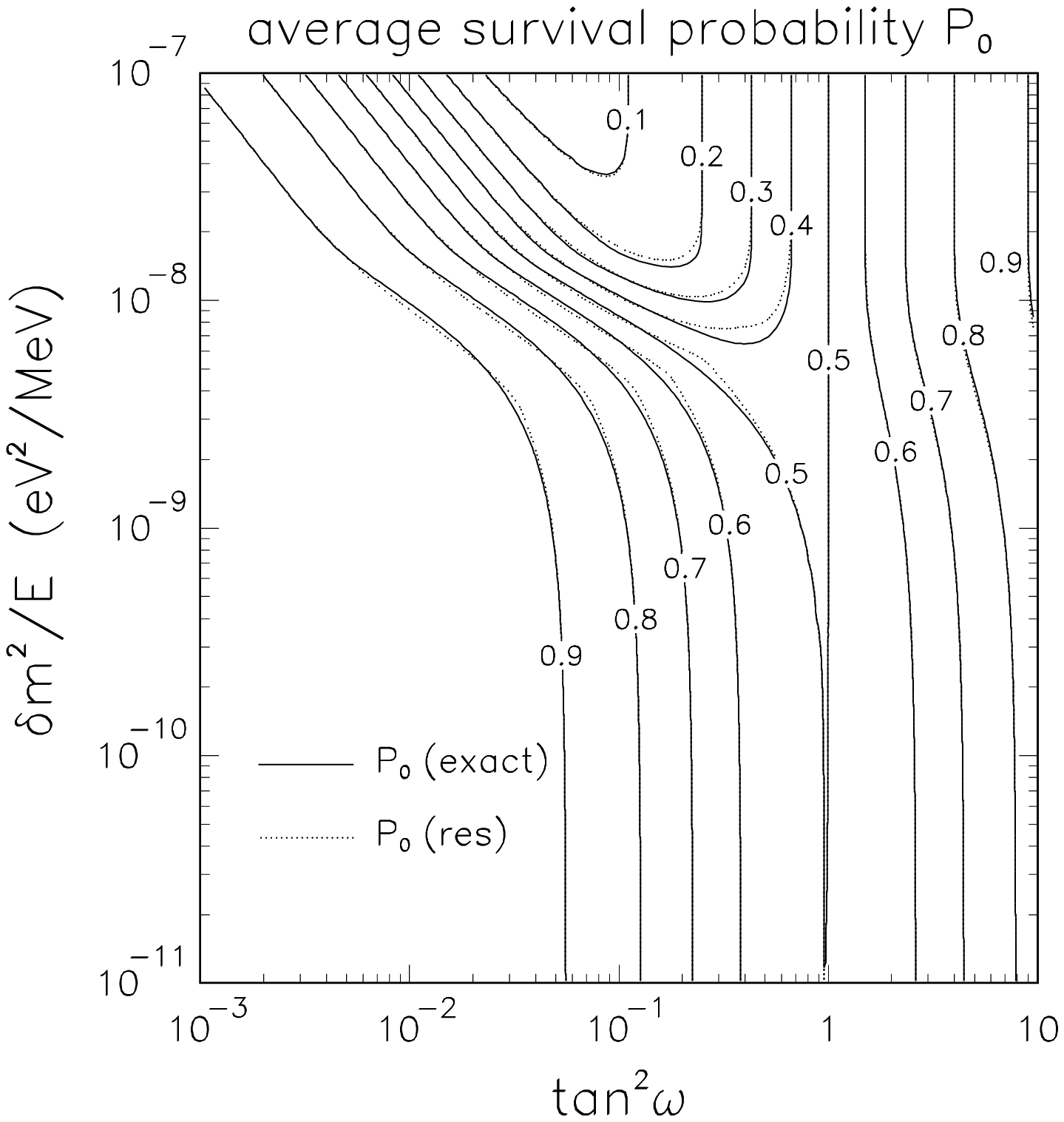}% 
{FIG.~2. The average survival probability $P_0$, as obtained
through exact numerical calculations (solid curves) and through the 
analytical resonance
prescription (dotted curves).
}
%..............................................................................
\InsertFigure{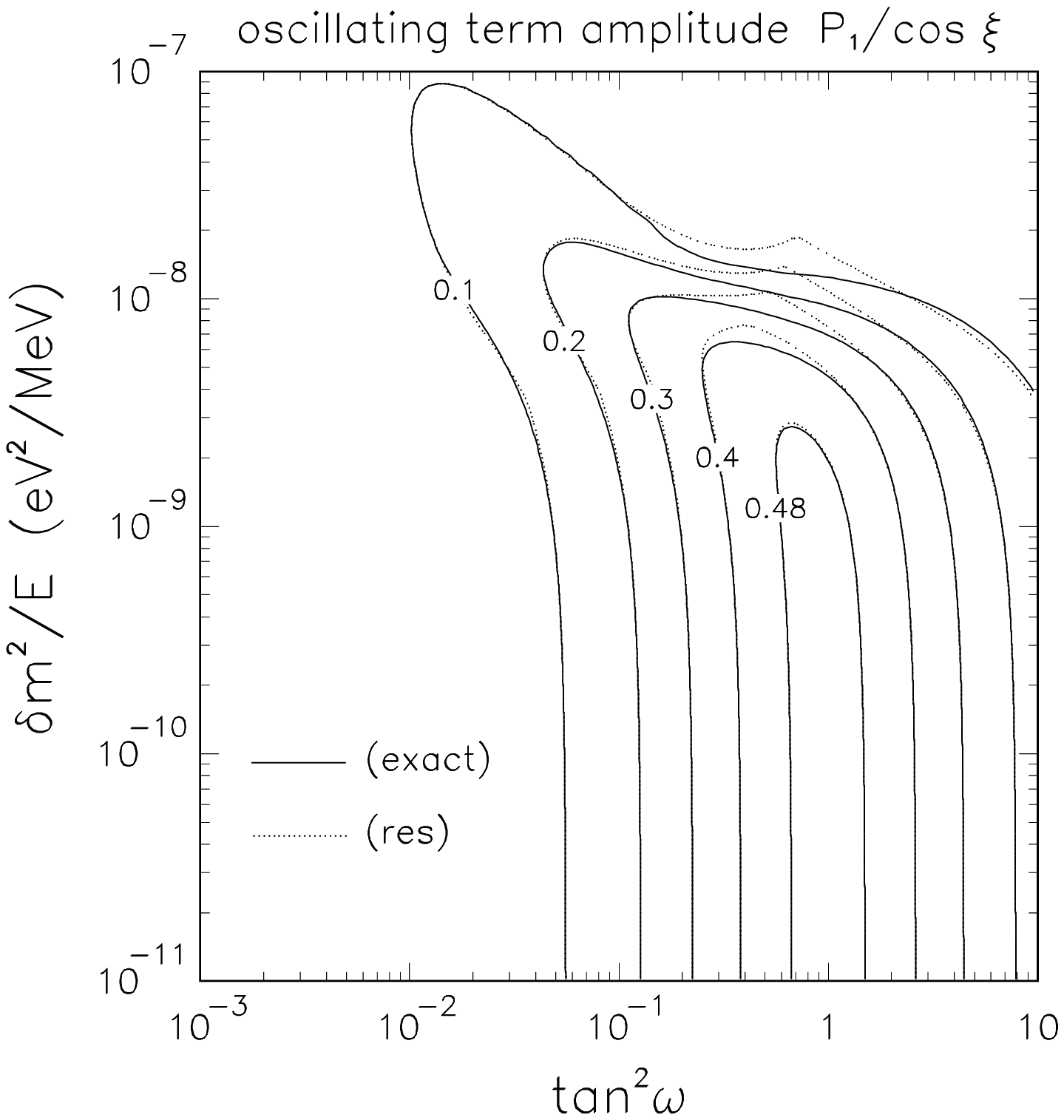}%
{FIG.~3. The oscillating term prefactor $P_1/\cos\xi$, as obtained
through exact numerical calculations (solid curves) and through the 
analytical resonance
prescription (dotted curves).
}
%..............................................................................
\InsertFigure{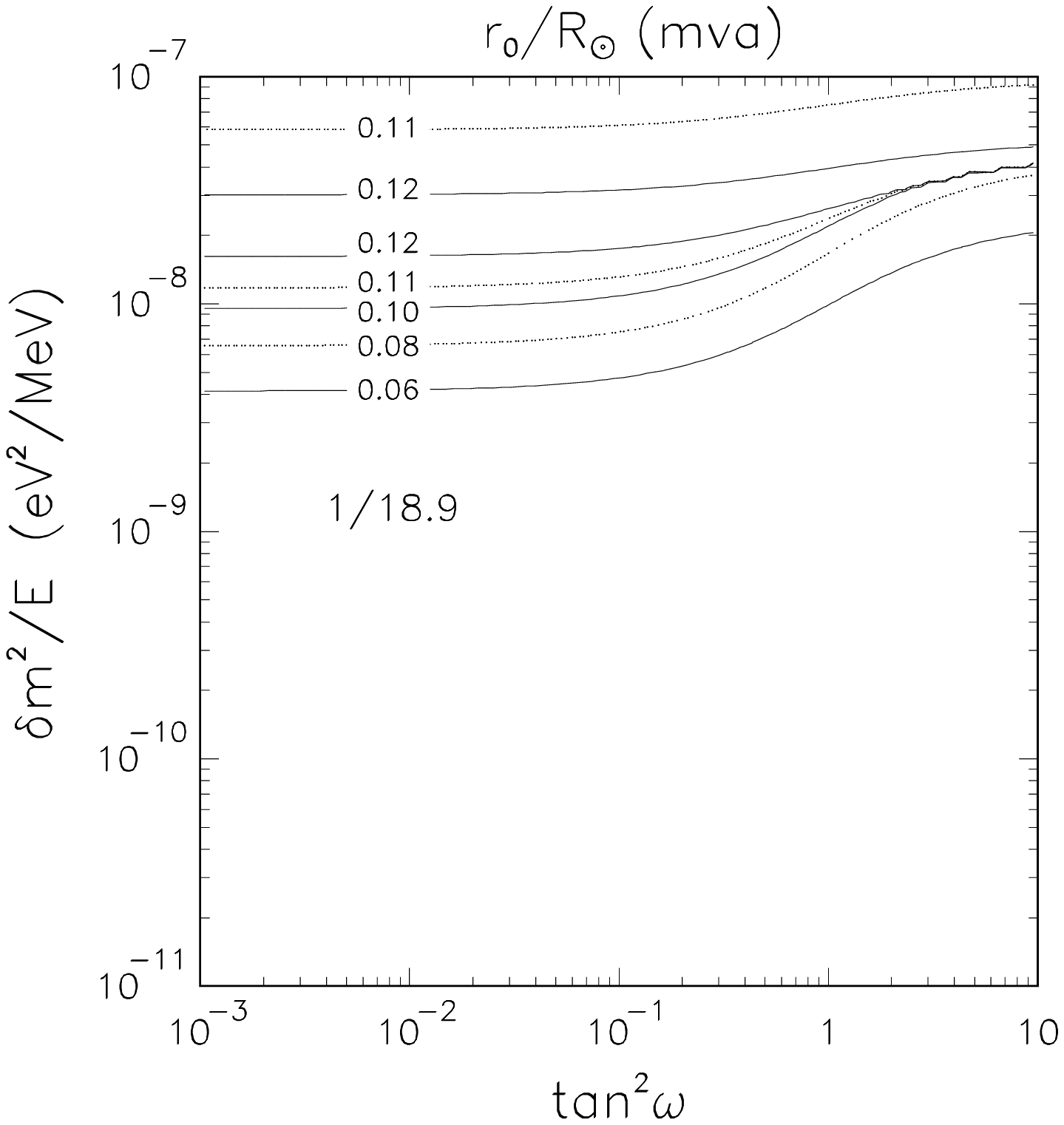}%
{FIG.~4. Isolines of the effective density scale parameter $r_0$ (in units
of $R_\odot$), as derived through the prescription of maximum violation
of adiabaticity (MVA), matched with the perturbative result valid at small
values of $\delta m^2/E$ ($r_0/R_\odot \to 1/18.9$).}
%..............................................................................
\InsertFigure{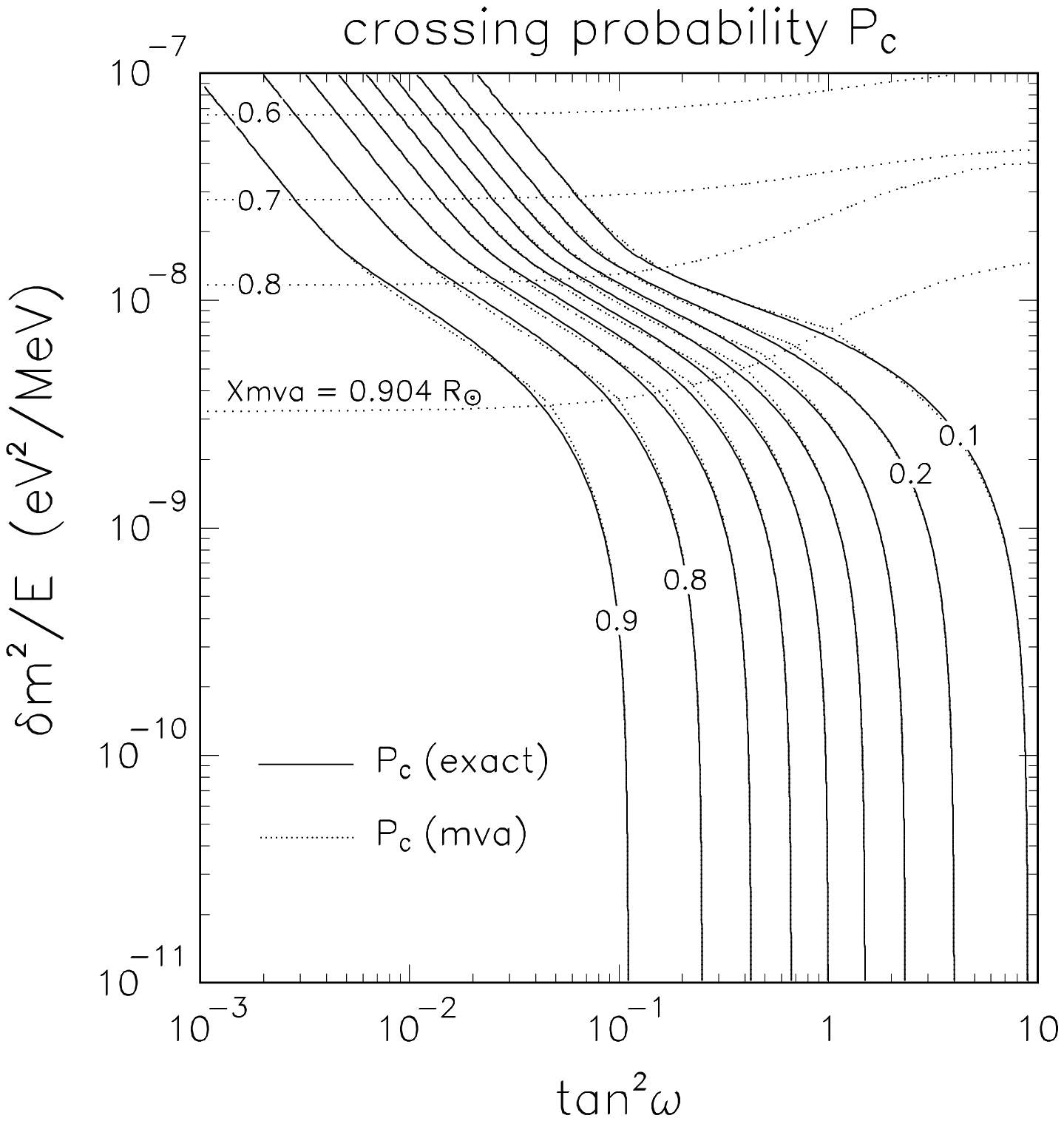}%
{FIG.~5.  The crossing probability $P_c$, as obtained through exact numerical
calculations (solid curves) and through the  analytical MVA prescription
(dotted curves). Also shown are isolines of MVA radii. The radius
$x/R_\odot=0.904$ represents the matching point of the MVA prescription with
the perturbative results valid at small values of $\delta m^2/E$.
}
%..............................................................................
\InsertFigure{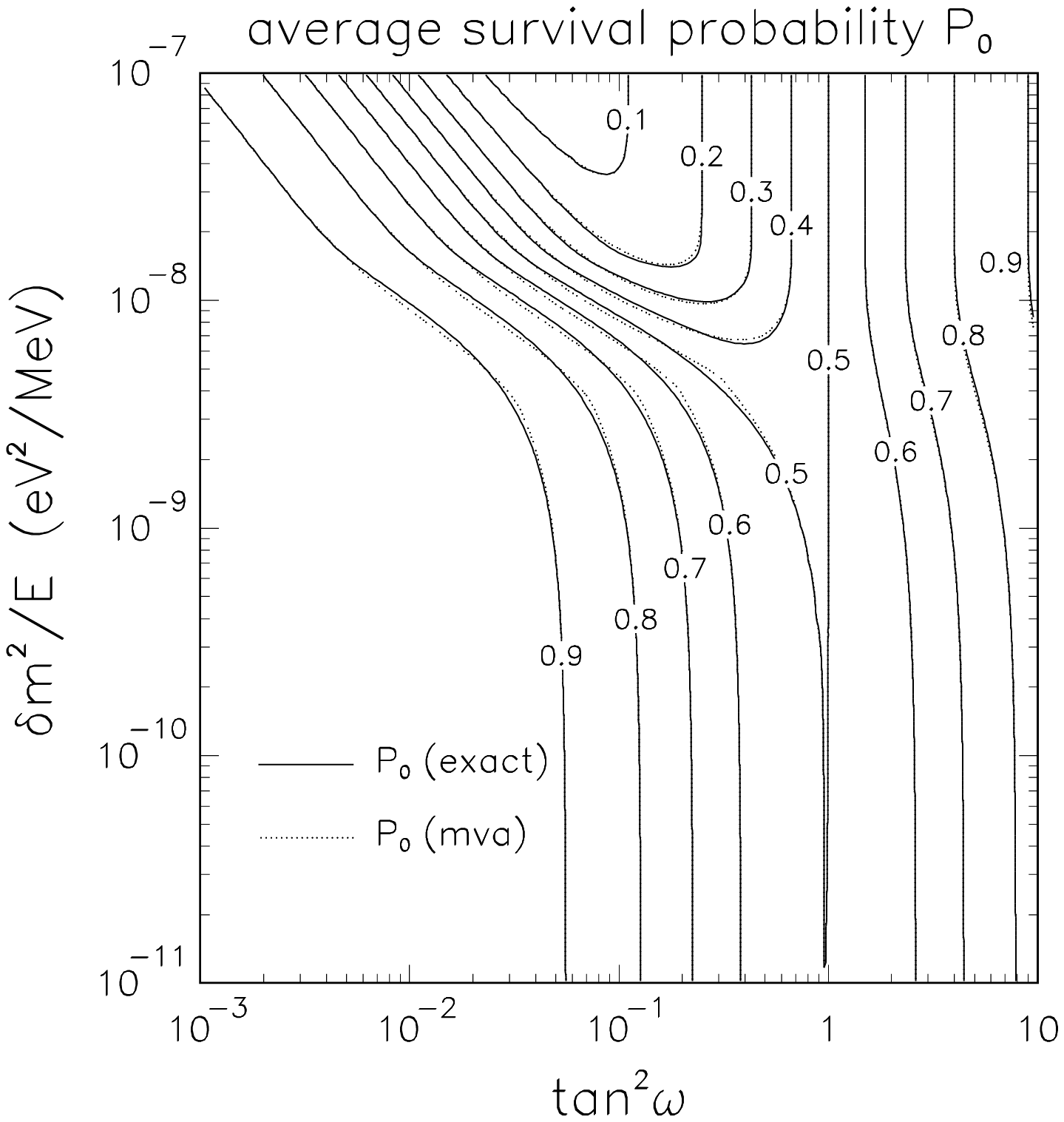}%
{FIG.~6. The average survival probability $P_0$, as obtained through exact
numerical calculations (solid curves) and through the  analytical MVA
prescription (dotted curves).
}
%..............................................................................
\InsertFigure{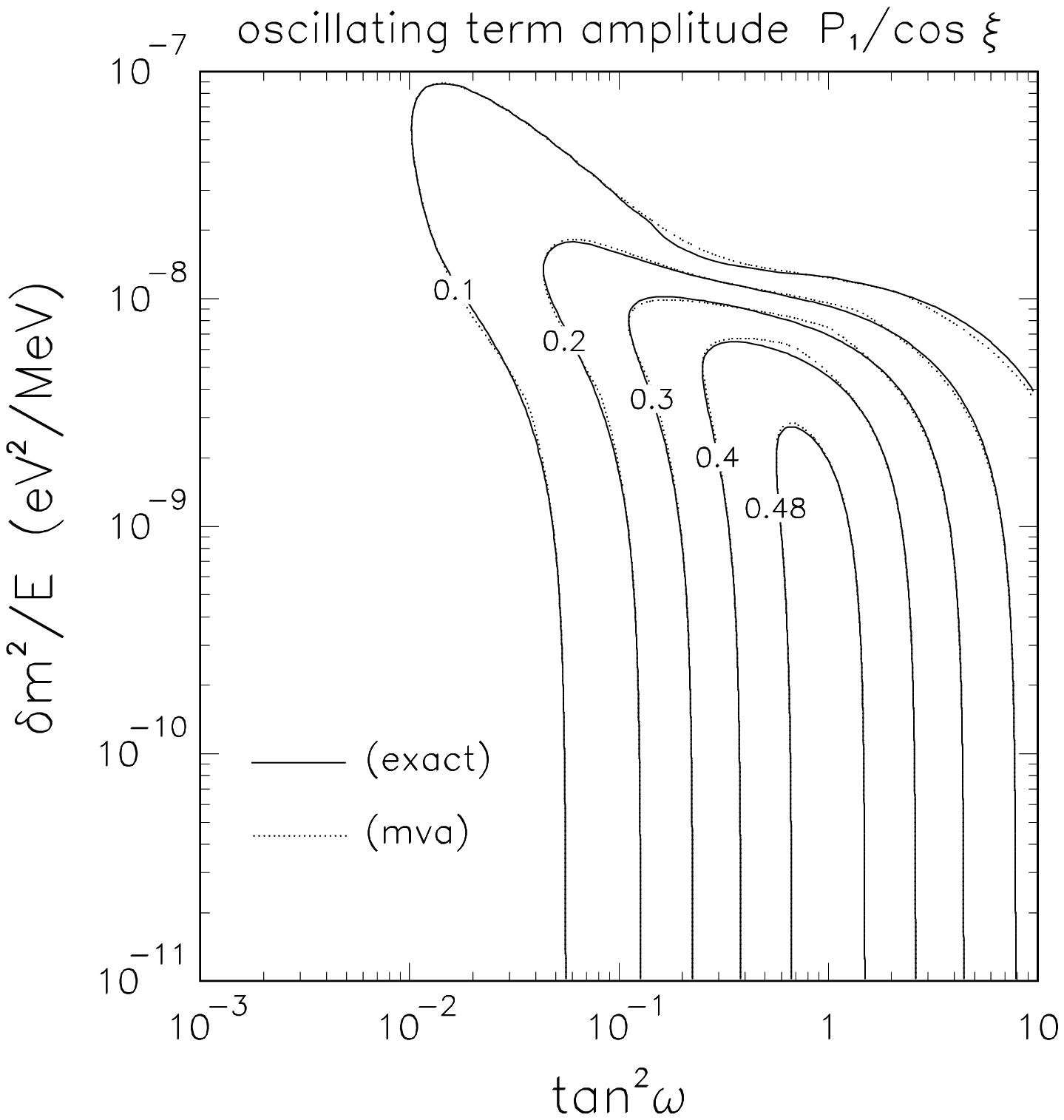}%
{FIG.~7. The oscillating term prefactor $P_1/\cos\xi$, as obtained through
exact numerical calculations (solid curves) and through the  analytical MVA
prescription (dotted curves).
}
%..............................................................................
\InsertFigure{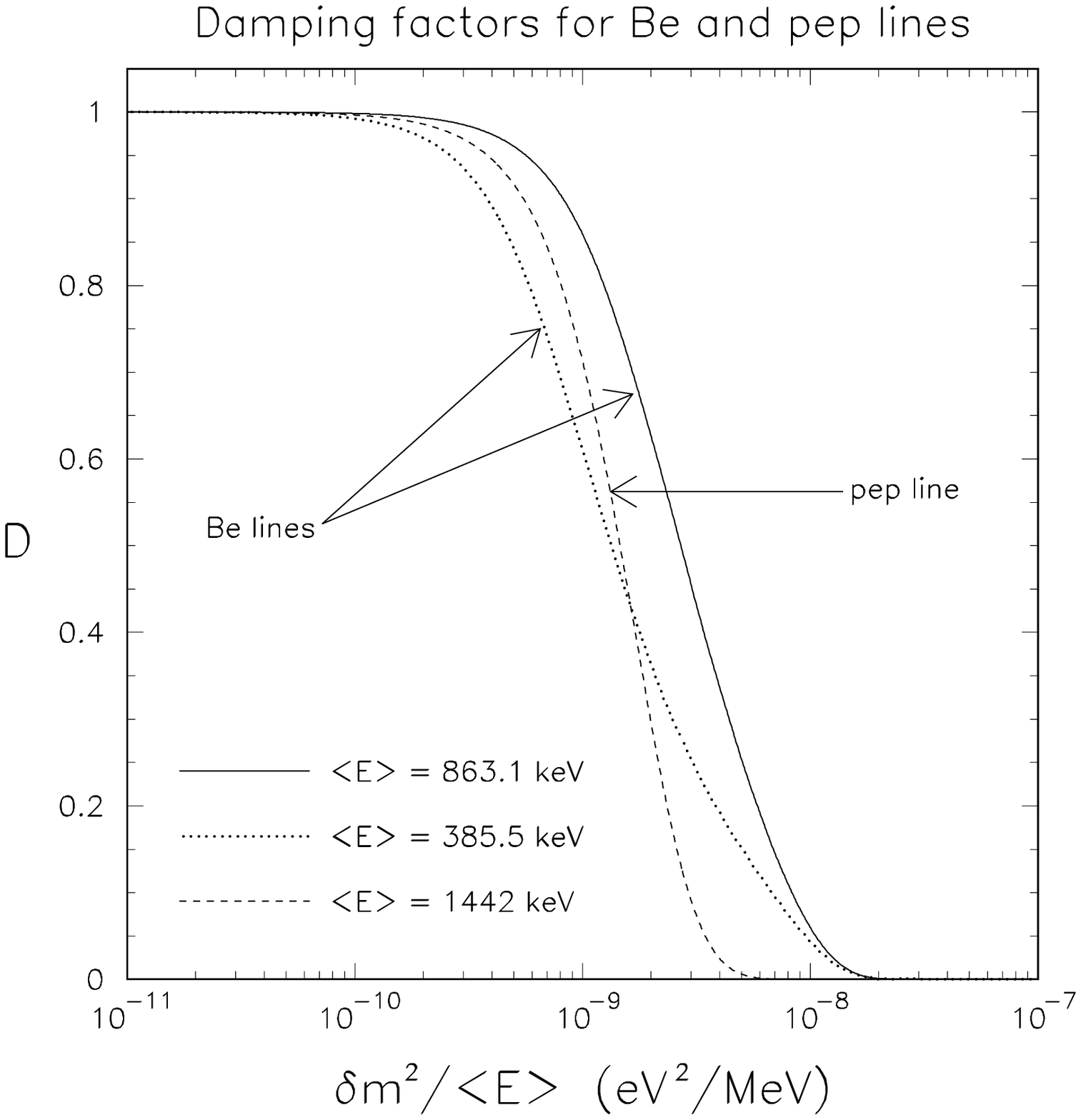}%
{FIG.~8. Damping factors for the oscillating term $\cos\xi$,
as calculated for the Be and {\em pep\/} solar neutrino lines 
($\langle E\rangle$ being their average energy). 
The damping factors completely suppress oscillations at the Earth for
$\delta m^2/\langle E \rangle$
above the quasivacuum range.}
%..............................................................................
\InsertFigure{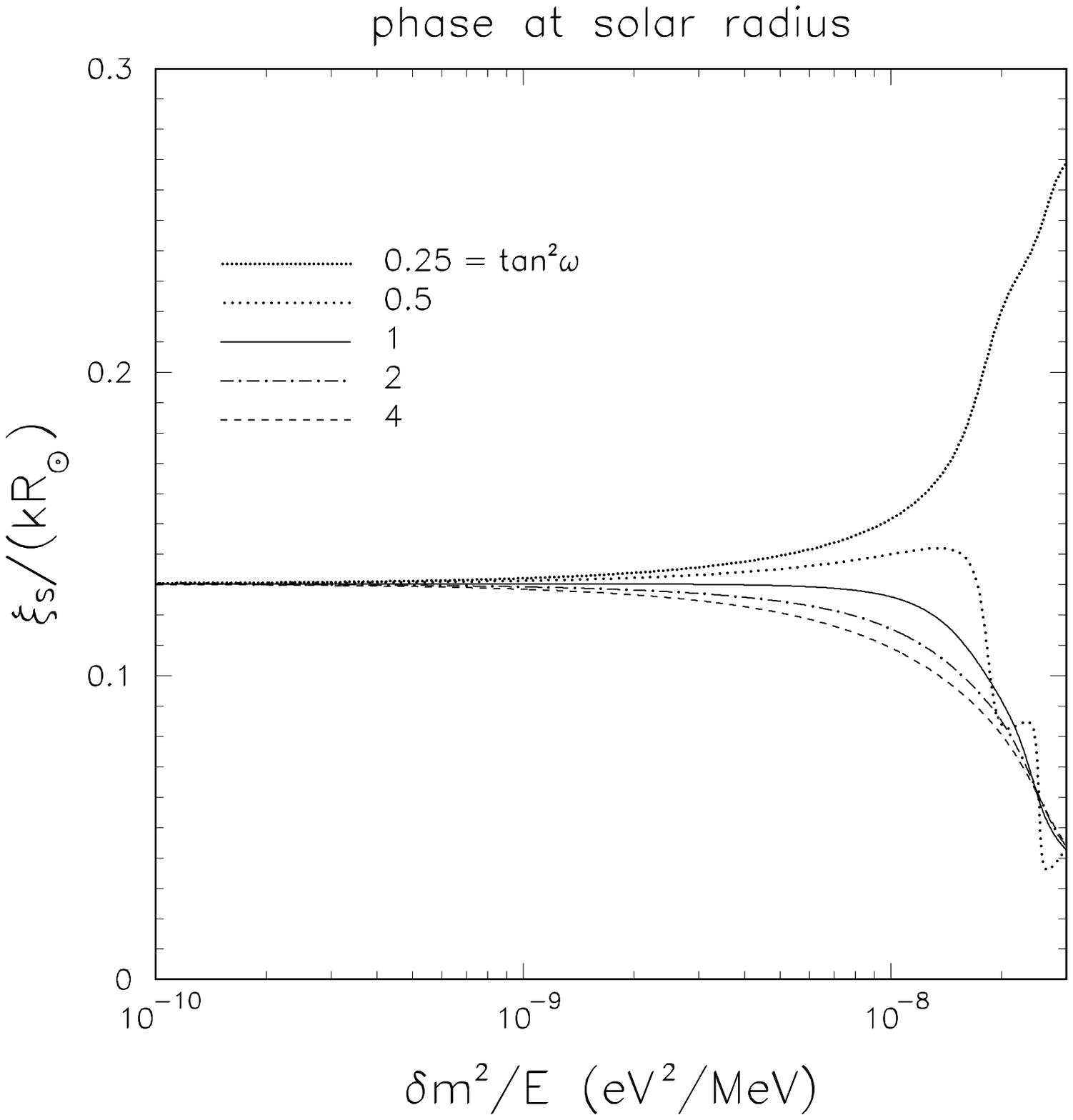}%
{FIG.~9. The solar phase $\xi_s$, in units of $kR_\odot$, as obtained from
exact numerical  calculations for some representative values of $\omega$. 
Notice the $\omega$-independent limit for $k\to0$.
}
%..............................................................................
\ShrinkFigure{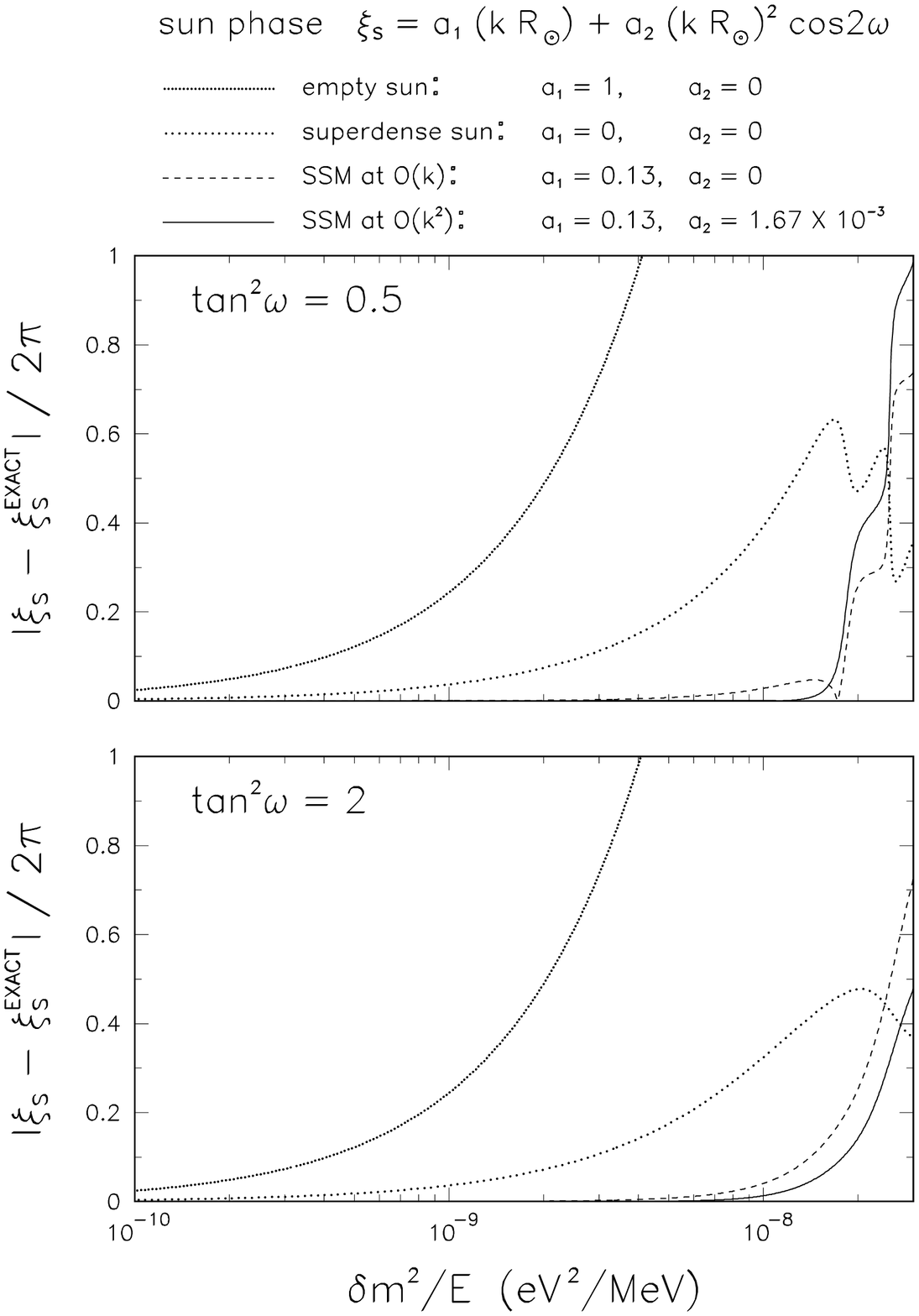}%
{FIG.~10. The absolute error induced by  approximate calculations
of the solar  phase $\xi_s$, as compared with the exact numerical calculations,
in units of $2\pi$. See the text for details.
}
%..............................................................................
\InsertFigure{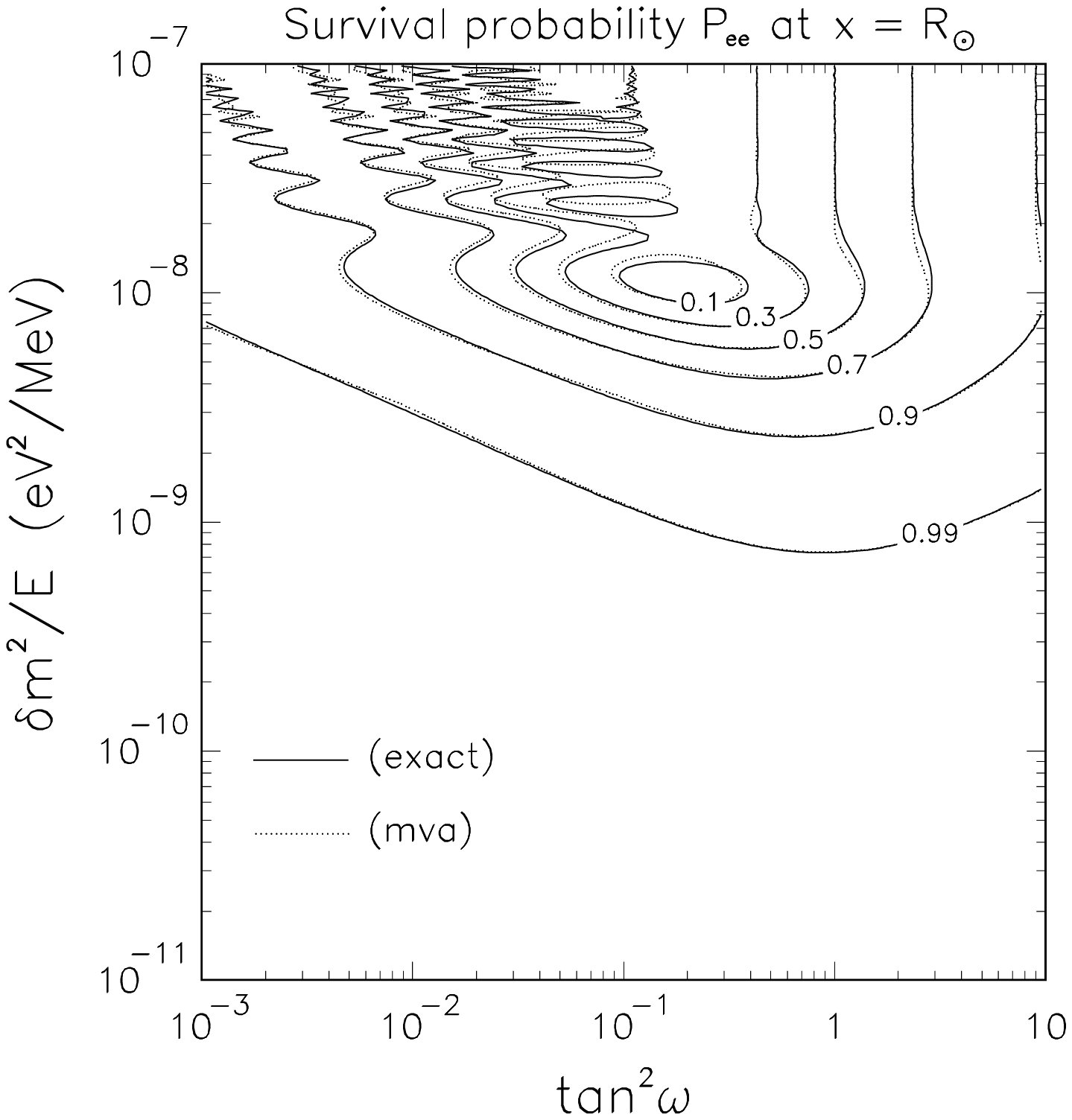}%
{FIG.~11. Application of our final analytical recipe for $P_c$ and $\xi_s$ to
the calculation of $P_{ee}(R_\odot)$  at the exit from the Sun (dotted lines),
as compared with the corresponding exact numerical results (solid lines).
}


\begin{thebibliography}{99}



\bibitem{NuAs}	J.N.\ Bahcall, 
		{\em Neutrino Astrophysics} (Cambridge University Press,
		Cambridge, England, 1989).

\bibitem{Cl98}	Homestake Collaboration, 
		B.T.\ Cleveland, T.J.\ Daily, R.\ Davis Jr., J.R.\
		Distel, K.\ Lande, C.\ K.\ Lee, P.S.\ Wildenhain, and J.\
		Ullman, 
		Astrophys.\ J.\ {\bf 496}, 505 (1998).

\bibitem{Fu96}	Kamiokande Collaboration, 
		Y.\ Fukuda {\em et al.},
		Phys.\ Rev.\ Lett.\ {\bf 77}, 1683 (1996).

\bibitem{Ab99}	SAGE Collaboration, 
		J.N.\ Abdurashitov {\em et al.},
		Phys.\ Rev.\ C {\bf 60}, 055801 (1999).

\bibitem{Ha99}	GALLEX Collaboration, 
		W.\ Hampel {\em et al.},
		Phys.\ Lett.\ B {\bf 447}, 127 (1999).

\bibitem{SK00}	SuperKamiokande Collaboration, 
		Y.\ Fukuda {\em et al.},
		Phys.\ Rev.\ Lett.\  {\bf 82}, 2430 (1999).

\bibitem{Su00}	SuperKamiokande Collaboration,
		talk by Y.\ Suzuki in {\em Neutrino~2000},
		19th International Conference on Neutrino
		Physics and Astrophysics (Sudbury, Canada, 2000),
		to appear in the Proceedings.

\bibitem{GNOC}	GNO Collaboration,
		M.\ Altmann {\em et al.},
		Phys.\ Lett.\ B {\bf 490}, 16 (2000).

\bibitem{BP00}	J.N.\ Bahcall, M.H.\ Pinsonneault, and S.\ Basu,
		astro-ph/0010346.

\bibitem{BaPi}	J.N.\ Bahcall homepage, 
		http://www.sns.ias.edu/$^\sim$jnb
		(Neutrino Software and Data).

\bibitem{Po67}	B.\ Pontecorvo, 
		Zh.\ Eksp.\ Teor.\ Fiz.\ {\bf 53}, 1717 (1967) 
		[Sov.\ Phys.\ JETP {\bf 26}, 984 (1968)]; 
		Z.\ Maki, M.\ Nakagawa, and S.\ Sakata, 
		Prog.\ Theor.\ Phys.\ {\bf 28}, 675 (1962);
		V.\ Gribov and B.\ Pontecorvo,
		Phys.\ Lett.\ B {\bf 28}, 493 (1969).

\bibitem{Wo78}	L.\ Wolfenstein, 
		Phys.\ Rev.\ D {\bf 17}, 2369 (1978).

\bibitem{Mi85}	S.P.\ Mikheyev and A.Yu.\ Smirnov, 
		Yad.\ Fiz.\ {\bf 42}, 1441 (1985) 
		[Sov.\ J.\ Nucl.\ Phys.\ {\bf 42}, 913 (1985)];
		Nuovo Cimento C {\bf 9}, 17 (1986).

\bibitem{Ba80}  V.\ Barger, S.\ Pakvasa, R.J.N.\ Phillips and K.\ Whisnant,
		Phys.\ Rev.\ D {\bf 22}, 2718 (1980).

\bibitem{Revi}	S.M.\ Bilenky, C.\ Giunti, and W.\ Grimus,
		Prog.\ Part.\ Nucl.\ Phys.\ {\bf 43}, 1 (1999).

\bibitem{KrSm}	J.N.\ Bahcall, P.I.\ Krastev, and A.Yu.\ Smirnov,
		Phys.\ Rev.\ D {\bf 58}, 096016 (1998).

\bibitem{MoPa}	Talks by D.\ Montanino and by A.\ Palazzo
		at {\em NOW~2000}, 2nd Europhysics Neutrino
		Oscillation Workshop (Conca Specchiulla, Italy, 2000),
		to appear in the Proceedings. Transparencies available
		at http://www.ba.infn.it/$^\sim$now2000

\bibitem{Vale}	M.C.\ Gonzalez-Garcia,
		M.\ Maltoni, C.\ Pe{\~n}a-Garay, and J.W.F.\ Valle,
		hep-ph/0009350.
		
\bibitem{BiPe}	S.M.\ Bilenky and S.T.\ Petcov,
		Rev.\ Mod.\ Phys.\ {\bf 59}, 671 (1987). 

\bibitem{Pe87}	S.T.\ Petcov, 
		Phys.\ Lett.\ B {\bf 200}, 373 (1988).

\bibitem{Pe88}	S.T.\ Petcov,
		Phys.\ Lett.\ B {\bf 214}, 139 (1988).

\bibitem{PeRi}	S.T.\ Petcov and J.\ Rich, 
		Phys.\ Lett.\ B {\bf 224}, 426 (1989).

\bibitem{Pa90}	J.\ Pantaleone,
		Phys.\ Lett.\ B {\bf 251}, 618 (1990).

\bibitem{Pk91}	S.\ Pakvasa and J.\ Pantaleone,
		Phys.\ Rev.\ Lett.\ {\bf 65}, 2479 (1990).

\bibitem{Fr00}	A.\ Friedland, Phys.\ Rev.\ Lett.\ {\bf 85}, 936 (2000).

\bibitem{Dark}	A.\ de Gouv{\^e}a, A.\ Friedland, and H.\ Murayama,
		Phys.\ Lett.\ B {\bf 490}, 125 (2000). 	

\bibitem{FoQV}	G.L.\ Fogli, E.\ Lisi, D.\ Montanino, and A.\ Palazzo,
		Phys.\ Rev.\ D {\bf 62}, 113004 (2000).

\bibitem{Spao}	A.M. Gago, H.\ Nunokawa, and R.\ Zukanovich Funchal,
		hep-ph/0007270.

\bibitem{Vosc}  See, e.g., 
		V.\ Barger, K.\ Whisnant, and R.J.N.\ Phillips,
		Phys.\ Rev.\ D {\bf 24}, 538 (1981);
		S.L.\ Glashow and L.M.\ Krauss, 
		Phys.\ Lett.\ B {\bf 190}, 199 (1987);
		A.\ Acker, S.\ Pakvasa, and J.\ Pantaleone,
		Phys.\ Rev.\ D {43}, 1754 (1991);
		P.I.\ Krastev and S.T.\ Petcov,
		Phys.\ Lett.\ B {\bf 285}, 85 (1992) and
                {\it ibid.} B {\bf 299}, 99 (1993);
		N.\ Hata and P.\ Langacker,
		Phys.\ Rev.\ D {\bf 56}, 6107 (1997); 
		B.\ Fa{\"\i}d, G.L.\ Fogli, E.\ Lisi, and D.\ Montanino,
		Astropart.\ Phys.\ {\bf 10}, 93 (1999).

\bibitem{Pe97}	S.T.\ Petcov, 
		Phys.\ Lett.\ B {\bf 406}, 355 (1997).

\bibitem{To87}	S.\ Toshev,
		Phys.\ Lett.\ B {\bf 196}, 170 (1987).
		
\bibitem{It88}	M.\ Ito, T.\ Kaneko, and M.\ Nakagawa,
		Prog.\ Theor.\ Phys.\ {\bf 79}, 13 (1988);
		{\em erratum\/} {\bf 79}, 555 (1988).

\bibitem{Ab92}  A.\ Abada and S.T.\ Petcov,
                Phys.\ Lett.\ B {\bf 279}, 153 (1992). 
 
\bibitem{Hax95} M.\ Bruggen, W.C.\ Haxton and Y.-Z.\ Quian,
                Phys.\ Rev.\ D {\bf 51}, 4028 (1995).   

\bibitem{Kr88}	P.I.\ Krastev and S.T.\ Petcov,
		Phys.\ Lett.\ B {\bf 207}, 64 (1988);
		{\em erratum}, {\bf 214}, 661 (1988).

\bibitem{FrQV}	A.\ Friedland,
		hep-ph/0010231.

\bibitem{Mu99}	A.\ de Gouvea, A.\ Friedland, and H.\ Murayama,
		hep-ph/9910286.

\bibitem{KuPa}	T.T.\ Kuo and J.\ Pantaleone,
		Rev.\ Mod.\ Phys.\ {\bf 61}, 937 (1989).

\bibitem{Fo96}	G.L.\ Fogli, E.\ Lisi, and D.\ Montanino,
		Phys.\ Rev.\ D {\bf 54}, 2048 (1996).

\bibitem{BU88}	J.N.\ Bahcall and R.\ Ulrich,
		Rev.\ Mod.\ Phys.\ {\bf 60}, 297 (1988).

\bibitem{MS87}	S.P.\ Mikheyev and A.Yu.\ Smirnov,
		in {\em Moriond~'87}, Proceedings of the 7th Moriond
		Workshop on New and Exotic Phenomena
		(Les Arcs, France, 1987), edited
		by O.\ Fackler and J.\ Tr{\^a}n Thanh V{\^a}n
		(Editions Fronti{\`e}res, Gif-sur-Yvette, France,
		1987), p.~405.

\bibitem{Ba86}	V.\ Barger, R.J.N.\ Phillips, and K.\ Whisnant,
		Phys.\ Rev.\ D {\bf 34}, 980 (1986).

\bibitem{Me86}	A.\ Messiah,
		in {\em Moriond~'86}, Proceedings of the 6th Moriond
		Workshop on Massive Neutrinos in Astrophysics and in
		Particle Physics (Tignes, France, 1986), edited
		by O.\ Fackler and J.\ Tr{\^a}n Thanh V{\^a}n
		(Editions Fronti{\`e}res, Gif-sur-Yvette, France,
		1986), p.~373.

\bibitem{Pa91}	J.\ Pantaleone, 
		Phys.\ Rev.\ D  {\bf 43}, 2636 (1991).

\bibitem{BaFr}	J.N.\ Bahcall and S.C.\ Frautschi,
		Phys.\ Lett.\ {\bf 29B}, 623 (1969).

\bibitem{BiPo}	S.M.\ Bilenky and B.\ Pontecorvo,
		Phys.\ Rep.\ {\bf 41}, 225 (1978). 

\bibitem{Cohe}	See A.S.\ Dighe, Q.Y.\ Liu, and A.Yu.\ Smirnov, 
		hep-ph/9903329, and references therein.

\bibitem{Line}	J.N.\ Bahcall,
		Phys.\ Rev.\ D {\bf 49}, 3923 (1994).

\bibitem{Kr85}	L.\ Krauss and F.\ Wilczek,
		Phys.\ Rev.\ Lett.\ {\bf 55}, 122 (1985).

\bibitem{BORE}	Borexino Collaboration, talk by G.\ Ranucci at
		{\em Neutrino~2000\/} \protect\cite{Su00}.

\bibitem{KAML}	KamLAND Collaboration, talk by A.\ Piepke at
		{\em Neutrino~2000\/} \protect\cite{Su00}.

\bibitem{Fr99}	A.\ de Gouv{\^e}a, A.\ Friedland, and H.\ Murayama,
		Phys.\ Rev.\ D {\bf 60}, 093011 (1999). 

\bibitem{BaWe}	A.J.\ Baltz and J.\ Weneser, Phys.\ Rev.\ D {\bf 37},
		3364 (1988).

\end{thebibliography}
\end{document}